\documentclass[12pt,a4paper]{paper}
\usepackage{lmodern}
\usepackage[margin=1in]{geometry}
\usepackage{fancyvrb}
\usepackage{amsmath,amssymb}
\usepackage{multirow}

\usepackage[font=small,labelfont=bf]{caption}
\allowdisplaybreaks
\usepackage{authblk}

\usepackage{multirow}
\newcommand{\bigzero}{\mbox{\normalfont\Large\bfseries 0}}
\newcommand{\bigI}{\mbox{\normalfont\Large\bfseries $I$}}
\usepackage[usenames,dvipsnames]{xcolor}
\usepackage{tikz}
\usepackage{subcaption}
\usetikzlibrary{tikzmark,decorations.pathreplacing,calc}
\usepackage{amsmath,amssymb}
\newcommand{\indep}{\raisebox{0.05em}{\rotatebox[origin=c]{90}{$\models$}}}

\usepackage{comment}

\newcommand{\Keywords}[1]{\def\@Keywords{#1}}
\let\code=\texttt
\let\proglang=\textsf
\newcommand{\pkg}[1]{{\fontseries{b}\selectfont #1}}

\newcommand{\doi}[1]{\href{http://dx.doi.org/#1}{\normalfont\texttt{\@doi{#1}}}}
\newcommand{\E}{\mathsf{E}}
\newcommand{\VAR}{\mathsf{VAR}}
\newcommand{\COV}{\mathsf{COV}}

\newcommand{\URL}[1]{\def\@URL{#1}}
\DefineVerbatimEnvironment{Code}{Verbatim}{}
\DefineVerbatimEnvironment{CodeInput}{Verbatim}{fontshape=sl}
\DefineVerbatimEnvironment{CodeOutput}{Verbatim}{}
\newenvironment{CodeChunk}{}{}

\usepackage{ragged2e}
\RaggedRight
\usepackage{hyperref}

\begin{document}
\title{mecor: An R package for measurement error correction in linear regression models with a continuous outcome}
\author[1,*]{Linda Nab}
\author[2]{Maarten van Smeden}
\author[3]{Ruth H. Keogh}
\author[1,4]{Rolf H.H. Groenwold}
\affil[1]{Department of Clinical Epidemiology, Leiden University Medical Center, Leiden, Netherlands}
\affil[2]{Julius Center for Health Sciences and Primary Care, University Medical Center Utrecht, Utrecht, Netherlands}
\affil[3]{Department of Medical Statistics,
London School of Hygiene and Tropical Medicine, London, United Kingdom}
\affil[4]{Department of Biomedical Data Sciences, Leiden University Medical Center, Leiden, the Netherlands}
\affil[*]{Corresponding author: Postzone C7-P,
P.O. Box 9600, 2300 RC Leiden,
the Netherlands, l.nab@lumc.nl, tel: +31 71 526 5640}
\maketitle 
\hrule
\begin{abstract}
  Measurement error in a covariate or the outcome of regression models is common, but is often ignored, even though measurement error can lead to substantial bias in the estimated covariate-outcome association. While several texts on measurement error correction methods are available, these methods remain seldomly applied. To improve the use of measurement error correction methodology, we developed \pkg{mecor}, an \proglang{R} package that implements measurement error correction methods for regression models with continuous outcomes. Measurement error correction requires information about the measurement error model and its parameters. This information can be obtained from four types of studies, used to estimate the parameters of the measurement error model: an internal validation study, a replicates study, a calibration study and an external validation study. In the package \pkg{mecor}, regression calibration methods and a maximum likelihood method are implemented to correct for measurement error in a continuous covariate in regression analyses. Additionally, methods of moments methods are implemented to correct for meaesurement error in the continuous outcome in regression analyses. Variance estimation of the corrected estimators is provided in closed form and using the bootstrap.
\end{abstract}
   \textit{Keywords}:~\@Keywords.
   \vskip 0.1in minus 0.05in
   \hrule
   \vskip 0.2in minus 0.1in


\section{Introduction}\label{sec:intro}
Measurement error is common across research fields, affecting the measurement of outcomes as well as important covariates. When left uncorrected, this can lead to severely biased and inefficient estimates of associations between covariates and outcome variables. Several texts have been published describing the impact of measurement error and, measurement error correction methodology
\cite{Buonaccorsi2010,Carroll2006,Fuller1987,Gustafson2004}. However, recent reviews by Brakenhoff et al. \cite{Brakenhoff2018} and Shaw et al. \cite{Shaw2018} show that, in biomedical research, measurement error correction methods remain seldomly applied. Keogh et al. \cite{KeoghSTRATOS:2020} suggest that one of the main barriers to the use of correction methods may be the lack of accessible software. Moreover, as exemplified in \cite{Wang2011}, measurement is not only common in biomedical research, but in bioinformatics, chemistry, astronomy and econometrics as well. Therefore, to facilitate and encourage the use of measurement error correction methodology, we developed \pkg{mecor}, an \proglang{R} package that provides measurement error correction methods for linear models with continuous outcomes. 

Several approaches to measurement error correction have been developed in the past decade. Examples include, simulation-extrapolation (SIMEX) by Cook et al. \cite{Cook1994Simex}, multiple imputation for measurement error by Cole et al. \cite{Cole2006MIME}, Bayesian correction (e.g., \cite{Bartlett2018BAY, Gustafson2004}), maximum likelihood-based methods (e.g., \cite{Bartlett2009LMM, Hesketh2003}), method of moments (MM) (e.g., \cite{Buonaccorsi2010}), and regression calibration (RC) introduced by Gleser \cite{Gleser1990RC} and Carroll et al. \cite{Carroll1990RC}. Of all these measurement error correction methods, RC is among the most commonly applied in biomedical research \cite{Shaw2018}, possibly because of its relative simplicity and the possibility to implement it in conjunction with a variety of analysis types, e.g., linear regression \cite{Gleser1990RC,Carroll1990RC}, survival analysis \cite{Prentice1982}), logistic regression \cite{Rosner1990} and other generalized linear models \cite{Armstrong1985, Carroll2006}.

In \proglang{R} \cite{R}, covariate measurement error correction by means of SIMEX is implemented in the package \pkg{simex} by Lederer et al. \cite{Lederer1997SIMEX}. The \proglang{R} package \pkg{simexaft} by He et al. \cite{He2012SIMEXaft} provides SIMEX covariate measurement error correction for accelerated failure time models. A special issue of the \proglang{Stata} \cite{Stata} Journal was published in 2003 and dedicated to measurement error models \cite{Sj2003}. Three different methods were introduced for correction of measurement error in covariates in a generalized linear model. The \code{rcal} and \code{eivreg} procedure were introduced for RC by Hardin et al. \cite{Hardin2003RC}, the \code{simex} and \code{simexplot} procedure were introduced for SIMEX by Hardin et al. \cite{Hardin2003SIMEX} and, the \code{cme} procedure was introduced by Rabe-Hesketh et al. \cite{Rabe-Hesketh2003} for measurement error correction using a maximum likelihood approach. In \proglang{SAS}, multiple macros have been developed for measurement error correction. These macros include \code{\%blinplus}, implementing the method by Rosner et al. \cite{Rosner1990}), \code{\%relibpls8}, implementing the method by Rosner et al. \cite{Rosner1992},  and \code{\%rrc}, implementing the method by Liao et al. \cite{Liao2011}), and the NCI method macros, implementing the methods by Kipnis et al. \cite{Kipnis2009}. An overview of available software including useful web links can be found in Table 4 and 5 of the paper by Keogh et al. \cite{KeoghSTRATOS:2020}. Although several measurement error correction methods are available in \proglang{Stata} and \proglang{SAS}, to date RC-like methods for measurement error correction in a covariate have not been implemented in an \proglang{R} package. Moreover, no method for measurement error correction in a continuous outcome has been implemented in \proglang{R}.

In this paper we present and describe \pkg{mecor}, an \proglang{R} package for measurement error correction in linear regression models with a continuous outcome. Several methods (i.e., RC, MM and maximum likelihood) are implemented to correct covariate-outcome associations for measurement error in a covariate, or in the outcome. The package \pkg{mecor} is flexible regarding the information that can be used to enable the measurement error correction, which can be of either of four types of measurement validation studies: an internal validation study, a replicates study, a calibration study and an external validation study. For each of these types of validation studies, standard RC, validation RC, efficient RC by Spiegelman et al. \cite{Spiegelman2001ERC} and a maximum likelihood approach by Bartlett et al. \cite{Bartlett2009LMM} are implemented for measurement error correction in a covariate. For outcome measurement error correction, standard MM \cite{Buonaccorsi2010} and efficient MM \cite{Keogh2016} are available, for all different types of validation studies except replicates studies. The package \pkg{mecor} allows for random or systematic measurement error in a covariate, systematic measurement error in the outcome and, additionally, differential outcome measurement error in a univariable analysis. This broad spectrum of validation study types, measurement error models and correction methods in our easy-to-use software package should improve the application of measurement error corrections in research practice.

This paper is organized as follows. Section 2 introduces several measurement error models and the data structures of the four validation study types that can be used to estimate the parameters of the measurement error model. Section 3 outlines the measurement error correction methods. Section 4 introduces the functions in the package \pkg{mecor}. Section 5 demonstrates how the package \pkg{mecor} can be used in different settings using simulated example data. 



\section{Measurement error: notation, types and data structures}\label{sec:memodel}
In this section, we introduce notation, derive expressions for the impact of measurement error on covariate-outcome associations and introduce the data structure of four different types of studies, that provide input for measurement error correction methods. Throughout, it is assumed that there is a continuous outcome $Y$, a continuous covariate $X$ and a vector of $k$ other covariates $\boldsymbol{Z}=(Z_1,Z_2,Z_3,\dots,Z_k)$. We consider measurement error in one variable at a time, i.e., in the covariate, $X$, or in the outcome, $Y$ and assume that the other variables in the model are measured without error. Since our focus is on studies in which we aim to estimate the covariate-outcome association, the covariate $X$ could be the main exposure of interest or a variable that confounds the relation between the main exposure and the outcome (one of the $Z$ variables). The parameters of interest are $\boldsymbol{\beta}=(\beta_X, \beta_0, \boldsymbol{\beta}_Z)$ (with $\boldsymbol{\beta}_Z$ a $1 \times k$ matrix) from the linear model,
\begin{equation}\label{eq:genmod}
    Y=\beta_X X + \beta_0 + \boldsymbol{\beta}_Z \boldsymbol{Z}' + e, \quad \mathrm{Var}(e) = \sigma^2,
\end{equation}
where we assume that $\E(e)=0$ and $\mathrm{Cov}(e,X)=\mathrm{Cov}(e,\boldsymbol{Z})=0$. This model will be referred to as the \textbf{outcome model}. 

\subsection{Types of measurement error and their impact}\label{sec:me:impact}
To quantify the impact of measurement error, we first define the assumed measurement error models. Subsequently, we outline the impact of measurement error in a covariate and the outcome on the estimates of the outcome model parameters, separately.

\subsubsection{Covariate measurement error} 
Let $X^*$ denote the error-prone substitute measure of the error-free reference measure $X$, following the measurement error model,
\begin{equation}\label{eq:covmem}
    X^* = \theta_0 + \theta_1 X + U, \quad \mathrm{Var}(U) = \tau^2,
\end{equation}
and assume that $\E(U)=0$ and $\mathrm{Cov}(U, X) = 0$. We assume non-differential covariate measurement error (i.e., $X^* \indep Y|X,\boldsymbol{Z}$ or, equivalently, that the errors $U$ are independent of the errors $e$ in equation (\ref{eq:genmod})). The measurement error is called `classical' or `random' if $\theta_0=0$ and $\theta_1=1$. The terms \textit{classical measurement error} and \textit{random measurement error} are used interchangeably in the literature. In this paper, we use the term random measurement error to refer to this type of measurement error. The measurement error is called `systematic' for all other values of $\theta_0$ and $\theta_1$.

Suppose that there is one covariate $\boldsymbol{Z} = Z_1$ in the outcome model in (\ref{eq:genmod}), and that data on $Y$, $X^*$ and $Z_1$ are available to fit the linear model,
\begin{equation}\label{eq:genmodme}
    \E(Y|X^*, Z_1)=\beta_X^*X^*+\beta_0^* + \boldsymbol{\beta}_Z^* Z_1.
\end{equation}
In this model, the least squares estimators $\boldsymbol{\hat{\beta}}^*=(\hat{\beta}_X^*, \hat{\beta}_0^*, \hat{\beta}_Z^*)$, are biased for $\boldsymbol{\beta}$, and consistent and unbiased estimators for $\boldsymbol{\beta\Lambda}$ where $\boldsymbol{\Lambda}$ is the $3 \times 3$ \textbf{calibration model matrix}: 

\begin{eqnarray*}
\boldsymbol{\Lambda} = \left(
\begin{array}{ccc}
{\lambda}_{X^*} & {\lambda}_0 & {\lambda}_{Z_1}\\
0 & 1 & 0 \\
0 & 0 & 1
\end{array}\right).
\end{eqnarray*}
A well-known special case of the calibration model matrix is the attenuation factor. In particular, when there is random measurement error in the substitute error-prone measure $X^*$, we have $\beta_X^* = \lambda_{X^*} \beta$, where $\lambda_{X^*}$ is called the attenuation factor \cite{Spearman1904} or regression dilution factor \cite{Frost2000, Hutcheon2010}. When there is more than one $\boldsymbol{Z}$ covariate in the outcome model defined by equation (\ref{eq:genmod}), the \textbf{calibration model matrix} generalizes to the following $(2+k) \times (2+k)$ matrix:
\begin{eqnarray}\label{eq:calmodmatrix}
\boldsymbol{\Lambda} = \left(
\begin{array}{ccc}
{\lambda}_{X^*} & {\lambda}_0 & \boldsymbol{{\lambda}_Z}\\
\multicolumn{1}{c}{\multirow{3}[2]{*}{\bigzero}} & \multicolumn{2}{c}{\multirow{3}[2]{*}{\bigI}}\\
& & \\
& &
\end{array}\right),
\end{eqnarray}
where $\boldsymbol{{\lambda}_Z}$ is a $1 \times k$ matrix, $\boldsymbol{0}$ is a $(1+k)\times1$ null matrix and $\boldsymbol{I}$ is a $(1+k)\times(1+k)$ identity matrix.

\subsubsection{Outcome measurement error}
Let $Y^*$ denote the error-prone substitute measure of the error-free reference measure $Y$, following the measurement error model,
\begin{equation}\label{eq:outmem}
    Y^{*} = \theta_0 + \theta_1 Y + U, \quad \mathrm{Var}(U) = \tau^2,
\end{equation}
and assume that $\E(U)=0$ and $\mathrm{Cov}(U, Y)=0$. We assume non-differential outcome measurement error (i.e., $Y^* \indep X|Y,\boldsymbol{Z}$ or, equivalently, that the errors $U$ are independent of the errors $e$ in equation (\ref{eq:genmod})), unless specified otherwise. Random and systematic outcome measurement error are defined analogously to random and systematic covariate measurement error, respectively \cite{Keogh2014Toolkit,Nab2018}.

Suppose, again, that there is one covariate $\boldsymbol{Z} = Z_1$ in the outcome model in (\ref{eq:genmod}) and that data on $Y^*$, $X$ and $Z_1$ are available to fit the linear model,
\begin{equation}\label{eq:genmodmeout}
    \mathrm{E}[Y^*|X, Z_1]=\beta_X^* X+\beta_0^*+\beta_Z^* Z_1.
\end{equation} 
If the measurement error in $Y^*$ is random, the least squares estimators $\boldsymbol{\hat{\beta}}^*=(\hat{\beta}_X^*, \hat{\beta}_0^*, \hat{\beta}_Z^*)$ are unbiased for $\boldsymbol{\beta}$. In contrast, if the error in $Y^*$ is systematic, the least squares estimators $\boldsymbol{\hat{\beta}}^*=(\hat{\beta}_X^*, \hat{\beta}_0^*, \hat{\beta}_Z^*)$ are biased for $\boldsymbol{\beta}$ \cite{Buonaccorsi2010, Keogh2016, Nab2018}. In order to identify consistent estimators for $\boldsymbol{\beta}$ by matrix multiplication, we add the integer 1 to the vector $\boldsymbol{\hat{\beta}}^*$. Then, $(\boldsymbol{\hat{\beta}}^*, 1)$ are consistent and unbiased estimators for $(\boldsymbol{\beta}, 1)\boldsymbol{\Theta}$ where $\boldsymbol{\Theta}$ is the $4 \times 4$ outcome \textbf{measurement error model matrix}:
\begin{eqnarray*}
\boldsymbol{{\Theta}} =
\left(
\begin{array}{cccc}
\theta_1 & 0 & 0 & 0\\
0 & \theta_1 & 0 & 0\\
0 & 0 & \theta_1 & 0\\
0 & \theta_0  & 0 & 1
\end{array}\right).
\end{eqnarray*}
When there is more than one $\boldsymbol{Z}$ covariate in the outcome model defined in equation (\ref{eq:genmod}), the calibration model matrix generalizes to the following $(2+k+1) \times (2+k+1)$ outcome \textbf{measurement error model matrix}:
\begin{eqnarray}\label{eq:outmematrix}
\boldsymbol{{\Theta}} &=& \begin{pmatrix}
\theta_1 & \hdots & \hdots & 0\\
\vdots & \ddots & & \vdots\\
\vdots & & \theta_1 & \vdots\\
0 & \theta_0 & \hdots & 1
\end{pmatrix},
\end{eqnarray}
where $\boldsymbol{\hat{\Theta}}$ contains all zero's except on the diagonal and the $(2+k+1,2)$th element.

\subsubsection{Differential outcome measurement error in univariable analyses}
We assume non-differential measurement error in the outcome in all but the following special case. Suppose exposure $X$ is binary (e.g., in a two-arm controlled randomised trial) and that there are no other covariates $\boldsymbol{Z}$ in the outcome model defined by equation (\ref{eq:genmod}). Further, suppose that the measurement error in $Y$ is differential such that the measurement error in the unexposed individuals (i.e., $X=0$) is different from the measurement error in the exposed individuals (i.e., $X=1$). Equivalently, let $Y^{*}$ be the error-prone substitute measure of the error-free reference measure $Y$, with mean $\E(Y^*|Y, X) = \theta_{X0} + \theta_{X1}Y$ and variance $\tau^2$, for $X = 0, 1$. Suppose now that data on $Y^*$ and $X$ are available to fit the linear model,
\begin{equation*}
    \mathrm{E}[Y^*|X]=\beta_X^* X+\beta_0^*.
\end{equation*} 

In this model, the least squares estimators $\boldsymbol{\hat{\beta}}^*=(\hat{\beta}_X^*, \hat{\beta}_0^*)$ are biased for $\boldsymbol{\beta}$ \cite{Keogh2016, Nab2018}. In order to identify consistent estimators for $\boldsymbol{\beta}$ by matrix multiplication, we again add the integer 1 to the vector $\boldsymbol{\hat{\beta}}^*$. Then, $(\boldsymbol{\hat{\beta}}^*,1)$ are consistent and unbiased estimators for $(\boldsymbol{\beta}, 1)\boldsymbol{\Theta}$ where, $\boldsymbol{\Theta}$ is the following $3 \times 3$ differential outcome \textit{measurement error model matrix}:
\begin{eqnarray}\label{eq:diffmematrix}
\boldsymbol{{\Theta}} = \begin{pmatrix}
{\theta}_{11} & 0 & 0 \\
{\theta}_{11}-{\theta}_{10} & {\theta}_{10} & 0\\
{\theta}_{01}-{\theta}_{00} & {\theta}_{00} & 1
\end{pmatrix}.
\end{eqnarray}
 
\subsection{Validation study data structures for measurement error correction}\label{sec:me:datastruc}
Four types of validation studies can be used to estimate the calibration model matrix or outcome measurement error model matrix defined in section \ref{sec:me:impact}: an internal validation study, a replicates study, a calibration study or an external validation study \cite{Keogh2019BookChapter,KeoghSTRATOS:2020}. The first three validation studies make use of information internal to the study cohort, whereas the fourth makes use of information external to the study cohort.

\subsubsection{Internal validation study}\label{sec:memodel:valstudy}
In an internal validation study, the error-free reference covariate values $X$ or outcome values $Y$ are observed in a subset of individuals (Table \ref{tab:valstudy}). Table \ref{tab:covvalstudy} shows the structure of an internal validation study for covariate measurement error. In the main study, the outcome $Y$, the error-prone substitute covariate $X^*$ and the covariates $\boldsymbol{Z}$ are measured in all $n$ individuals. Additionally, in $n_{\mathrm{sub}}$ individuals ($n_{\mathrm{sub}}<n$) the true covariate $X$ is measured, assumed a random subset of the main study. As an example, suppose the true exposure of interest is visceral adipose tissue measurements (i.e., $X$) but that this is too expensive to obtain on all study participants and the error-prone substitute measure of waist circumference is instead collected for everyone (i.e, $X^*$). The same structure holds for an internal validation study for outcome measurement error, as shown in Table \ref{tab:outvalstudy}. 

\begin{table}[!htb]
    \caption{\textbf{Data structure of internal validation studies.} The true covariate or outcome is observed in a subset of the individuals from the main study. The superscript $*$ indicates that the variable was measured with error.}\label{tab:valstudy}
    \begin{subtable}{.5\linewidth}
      \centering
        \caption{Covariate-validation study}\label{tab:covvalstudy}
    \begin{tabular}{c c c c} 
      $Y$ & $X^*$ & $\boldsymbol{Z}$ & $X$ \\
      \hline
      $y_1$  & $x_1^*$ & $\boldsymbol{z_1}$ & $x_1$ \\
      \vdots & \vdots  & \vdots & \vdots\\
      \vdots & \vdots  & \vdots &$x_{n_{\mathrm{sub}}}$\\
      \vdots & \vdots  & \vdots & - \\ 
      \vdots & \vdots  & \vdots & \vdots \\
      $y_n$  & $x_n^*$ & $\boldsymbol{z_n}$ & - \\
    \end{tabular}
    \end{subtable}
    \begin{subtable}{.5\linewidth}
      \centering
        \caption{Outcome-validation study}\label{tab:outvalstudy}
    \begin{tabular}{c c c c} 
      $Y^*$ & $X$ & $\boldsymbol{Z}$ & $Y$ \\
      \hline
      $y^*_1$  & $x_1$ & $\boldsymbol{z_1}$ & $y_1$ \\
      \vdots & \vdots  & \vdots & \vdots\\
      \vdots & \vdots  & \vdots &$y_{n_{\mathrm{sub}}}$\\
      \vdots & \vdots  & \vdots & - \\ 
      \vdots & \vdots  & \vdots & \vdots \\
      $y^*_n$  & $x_n$ & $\boldsymbol{z_n}$ & - \\
    \end{tabular}
    \end{subtable} 
\end{table}

\subsubsection*{Replicates study}
A replicates study can be used if the measurement error in a covariate is random, denoted by $X^{*_r}$. We will only use this type of study for covariate measurement error since random measurement error in an outcome does not result in biased association estimates (section \ref{sec:me:impact}). In a replicates study, the error-prone substitute covariate $X^{*_r}$ is repeatedly measured (i.e., $m$ times, where $m \geq 2$) in all or in a random subset of individuals (Table \ref{tab:repstudy}). The repeated measures are denoted by $X_1^{*_r}, \dots, X_m^{*_r}$. We assume that, in each individual, the same number of repeated measures was observed. Further, we assume that the measurement error in the replicates is jointly independent. Table \ref{tab:repstudy:wide} and \ref{tab:repstudy:long} show the structure of a replicates study with full and partial replicates, respectively. In the main study, the outcome $Y$, the error-prone substitute covariate $X_1^{*_r}$ and the covariates $\boldsymbol{Z}$ are measured in all $n$ individuals. Additionally, $n_{\mathrm{sub}} \leq n$ individuals have $m$ replicates of the error-prone substitute measure $X_j^{*_r}$ for $j = 2 \dots m$. An example is the repeated measurement of several coronary risk factors in the Framingham Heart study, such as serum cholesterol, blood glucose, and systolic blood pressure \cite{Rosner1992}. 

\begin{table}[!htb]
    \caption{\textbf{Data structure of a covariate-replicates study for full or partial replicates.} The error-prone covariate is measured $m$ times in all or a subset of individuals. The superscript $*_r$ indicates random measurement error.}\label{tab:repstudy}
    \begin{subtable}{.5\linewidth}
      \centering
        \caption{Full replicates study}\label{tab:repstudy:wide}
      \begin{tabular}{c c c c c c} 
      $Y$ & $X_1^{*_r}$ & $\boldsymbol{Z}$ & $X_2^{*_r}$ & $\hdots$ & $X^{*_r}_m$ \\
      \hline
      $y_1$  & $x_{11}^{*_r}$  & $\boldsymbol{z_1}$ & $x^{*_r}_{12}$ & $\hdots$ & $x^{*_r}_{1m}$ \\
      \vdots & \vdots & \vdots & \vdots & \vdots & \vdots \\
      \vdots & \vdots & \vdots & \vdots & \vdots & \vdots \\
      \vdots & \vdots & \vdots & \vdots & \vdots & \vdots \\
      \vdots & \vdots & \vdots & \vdots & \vdots & \vdots \\
      $y_n$  & $x_{n1}^{*_r}$ & $\boldsymbol{z_n}$ & $x^{*_r}_{n2}$ & $\hdots$ & $x^{*_r}_{nm}$\\
    \end{tabular}
    \end{subtable}%
    \begin{subtable}{.5\linewidth}
      \centering
        \caption{Partial replicates study}\label{tab:repstudy:long}
                  \begin{tabular}{c c c c c c} 
      $Y$ & $X_1^{*_r}$ & $\boldsymbol{Z}$ & $X_2^{*_r}$ & $\hdots$ & $X^{*_r}_m$ \\
      \hline
      $y_1$  & $x_{11}^{*_r}$  & $\boldsymbol{z_1}$ & $x^{*_r}_{12}$ & $\hdots$ & $x^{*_r}_{1m}$ \\
      \vdots & \vdots & \vdots & \vdots & $\hdots$ & \vdots \\
      \vdots & \vdots & \vdots & $x^{*_r}_{n_{\mathrm{sub}}2}$ & $\hdots$ & $x^{*_r}_{n_{\mathrm{sub}}m}$ \\
      \vdots & \vdots & \vdots & - & $\hdots$ & -\\ 
      \vdots & \vdots & \vdots & \vdots & \vdots & \vdots \\
      $y_n$  & $x_{n1}^{*_r}$ & $\boldsymbol{z_n}$ & - & $\hdots$ & -\\
    \end{tabular}
    \end{subtable} 
\end{table}

\subsubsection*{Calibration study} 
A calibration study is a special type of sub-study where two types of error-prone substitute measurement methods are used to measure the covariate or outcome: substitute measurement prone to systematic measurement error and a substitute measurement prone to random measurement error (Table \ref{tab:calstudy}). Table \ref{tab:covcalstudy} shows the structure of a calibration study for covariate measurement error. All $n$ individuals in the main study have obtained measures of the outcome $Y$, the error-prone substitute covariate $X^{*_s}$ and the covariates $\boldsymbol{Z}$. The error-prone substitute covariate $X^{*_s}$ is systematically different from $X$, or, $\E(X^{*_s}|X)\neq X$ (systematic measurement error). Additionally, a random subset of $n_{\mathrm{sub}}$ individuals ($n_{\mathrm{sub}}< n$) have $m$ replicates of the error-prone substitute measure $X_j^{*_r}$, where $\E(X_j^{*_r}|X)= X$ for $j = 1 \dots m$ (random measurement error). The same structure holds for a calibration study for outcome measurement error, as shown in Table \ref{tab:outcalstudy}. An example of an calibration study for outcome measurement error is a study of sodium intake measured by a 24-hour recall (assumed systematic measurement error) and urinary biomarkers (assumed random measurement error) \cite{Keogh2016}.

\begin{table}[!htb]
    \caption{\textbf{Data structure of calibration studies.} Two types of error-prone measurement methods are used to measure the covariate or outcome. The superscripts $*_r$ and $*_s$ indicate random and systematic measurement error, respectively.}\label{tab:calstudy}
    \begin{subtable}{.5\linewidth}
      \centering
        \caption{Covariate-calibration study}\label{tab:covcalstudy}
    \begin{tabular}{c c c c c c} 
      $Y$ & $X^{*_s}$ & $\boldsymbol{Z}$ & $X_1^{*_r}$ & $\hdots$ & $X^{*_r}_m$ \\
      \hline
      $y_1$  & $x_{1}^{*_s}$  & $\boldsymbol{z_1}$ & $x^{*_r}_{11}$ & $\hdots$ & $x^{*_r}_{1m}$ \\
      \vdots & \vdots & \vdots & \vdots & $\hdots$ & \vdots \\
      \vdots & \vdots & \vdots & $x^{*_r}_{n_{\mathrm{sub}}1}$ & $\hdots$ & $x^{*_r}_{n_{\mathrm{sub}}m}$ \\
      \vdots & \vdots & \vdots & - & $\hdots$ & -\\ 
      \vdots & \vdots & \vdots & \vdots & $\vdots$ & \vdots \\
      $y_n$  & $x_{n}^{*_s}$ & $\boldsymbol{z_n}$ & - & $\hdots$ & -\\
    \end{tabular}
    \end{subtable}%
    \begin{subtable}{.5\linewidth}
      \centering
        \caption{Outcome-calibration study}\label{tab:outcalstudy}
            \begin{tabular}{c c c c c c} 
      $Y^{*_s}$ & $X$ & $\boldsymbol{Z}$ & $Y_1^{*_r}$ & $\hdots$ & $Y^{*_r}_m$ \\
      \hline
      $y^{*_s}_1$  & $x_{1}$  & $\boldsymbol{z_1}$ & $y^{*_r}_{11}$ & $\hdots$ & $y^{*_r}_{1m}$ \\
      \vdots & \vdots & \vdots & \vdots & $\hdots$ & \vdots \\
      \vdots & \vdots & \vdots & $y^{*_r}_{n_{\mathrm{sub}}1}$ & $\hdots$ & $y^{*_r}_{n_{\mathrm{sub}}m}$ \\
      \vdots & \vdots & \vdots & - & $\hdots$ & -\\ 
      \vdots & \vdots & \vdots & \vdots & $\vdots$ & \vdots \\
      $y^{*_s}_n$  & $x_{n}$ & $\boldsymbol{z_n}$ & - & $\hdots$ & -\\
    \end{tabular}
    \end{subtable} 
\end{table}

\subsubsection*{External validation study}\label{sec:memodel:exvalstudy}
In an external validation study the error-free reference covariate values $X$ or outcome values $Y$ are observed in a small set of individuals not included in the main study (Table \ref{tab:exvalstudy}). Table \ref{tab:exvalstudy:cov} shows the structure of an external validation study for covariate measurement error. In all $n$ individuals in the main study measures are obtained of outcome $Y$, the error-prone substitute covariate $X^*$ and the covariates $\boldsymbol{Z}$. Additionally, there is an external data set comprising of individuals on whom measures are obtained of the error-free reference covariate $X$, the error-prone substitute covariate $X^*$ and the other covariates $\boldsymbol{Z}$. Table \ref{tab:exvalstudy:out} shows the structure of an external validation study for outcome measurement error. In this setting, there is an external data set comprising of individuals of whom measures are obtained of the error-free reference outcome $Y$ and the error-prone substitute outcome $Y^*$. The external data set does not need to comprise measures of the covariates. An example of an external validation study for outcome measurement error is a trial designed to study the efficacy of iron supplementation in pregnant women where haemoglobin is measured in capillary blood samples (error-prone substitute measure) instead of in venous blood samples (error-free reference measure) \cite{Nab2018}.

\begin{table}[!htb]
    \caption{\textbf{Data structure of external validation studies.} An error-prone covariate or outcome is measured in the main study and the true covariate or outcome is measured in a small external set. The superscript $*$ indicates that there is random or systematic measurement error in the variables}\label{tab:exvalstudy}
    \begin{subtable}[t]{.5\linewidth}
      \centering
      \caption{External covariate-validation study}\label{tab:exvalstudy:cov}
        \begin{subtable}[t]{.40\linewidth}
        \centering
            \begin{tabular}[t]{c c c} 
            $Y$ & $X^*$ & $\boldsymbol{Z}$ \\
            \hline
            $y_1$  & $x_1^*$ & $\boldsymbol{z_1}$\\
            \vdots & \vdots  & \vdots\\
            \vdots & \vdots  & \vdots\\
            $y_n$  & $x_n^*$ & $\boldsymbol{z_n}$\\
            \multicolumn{3}{c}{{$\underbrace {\hspace{6em}}$}}\\
            \multicolumn{3}{c}{{Main study}}
            \end{tabular}
        \end{subtable}%
        \begin{subtable}[t]{.40\linewidth}
        \centering
            \begin{tabular}[t]{c c c} 
            $X$ & $X^*$ & $\boldsymbol{Z}$ \\
            \hline
            $x_1$  & $x_1^*$ & $\boldsymbol{z_1}$\\
            \vdots & \vdots  & \vdots\\
            $x_{n_{\text{ex}}}$  & $x_{n_{\text{ex}}}^*$ & $\boldsymbol{z_{n_{\text{ex}}}}$\\
            \multicolumn{3}{c}{{$\underbrace {\hspace{7em}}$}}\\
            \multicolumn{3}{c}{{\textit{External}}}
            \end{tabular}
        \end{subtable}%
    \end{subtable}%
    \begin{subtable}[t]{.5\linewidth}
    \centering
      \caption{External outcome-validation study}\label{tab:exvalstudy:out}
        \begin{subtable}[t]{.4\linewidth}
        \centering
            \begin{tabular}[t]{c c c} 
            $Y^*$ & $X$ & $\boldsymbol{Z}$ \\
            \hline
            $y_1^*$  & $x_1$ & $\boldsymbol{z_1}$\\
            \vdots & \vdots  & \vdots\\
            \vdots & \vdots  & \vdots\\
            $y_n^*$  & $x_n$ & $\boldsymbol{z_n}$\\
            \multicolumn{3}{c}{{$\underbrace {\hspace{6em}}$}}\\
            \multicolumn{3}{c}{{Main study}}
            \end{tabular}
        \end{subtable}%
        \begin{subtable}[t]{.3\linewidth}
        \centering
            \begin{tabular}[t]{c c} 
            $Y$ & $Y^*$  \\
            \hline
            $y_1$  & $y_1^*$ \\
            \vdots & \vdots \\
            $y_{n_{\text{ex}}}$  & $y_{n_{\text{ex}}}^*$ \\
            \multicolumn{2}{c}{{$\underbrace {\hspace{5em}}$}}\\
            \multicolumn{2}{c}{{\textit{External}}}
            \end{tabular}
        \end{subtable}%
    \end{subtable}
\end{table}

\section{Measurement error correction}\label{sec:meascor}
In section \ref{sec:me:impact}, the calibration model matrix $\boldsymbol{\Lambda}$ and the measurement error model matrix $\boldsymbol{\Theta}$ were introduced. These matrices quantify the bias in the naive analysis, i.e., the analysis that does not take the measurement error in $X^*$ or $Y^*$ into account. In the following sections, measurement error correction methods are introduced that utilize the matrices $\boldsymbol{\Lambda}$ and $\boldsymbol{\Theta}$. 

The standard method for covariate measurement error correction that uses the calibration model matrix $\boldsymbol{\Lambda}$ is \textit{standard regression calibration (RC)} \cite{Gleser1990RC,Carroll1990RC}. \textit{Standard RC} can be applied in all four types of studies from the previous section. In addition, \textit{validation RC}, an adapted version of \textit{standard RC} for internal validation studies, is the standard covariate measurement error correction method for internal validation studies \cite{Carroll2006}. Further, the standard method for outcome measurement error correction that uses the measurement error model matrix $\boldsymbol{\Theta}$ is \textit{standard method of moments (MM)} \cite{Buonaccorsi2010}. \textit{Standard MM} can be applied in internal and external validation studies, and calibration studies. 

\textit{Standard RC} and \textit{standard MM} do not make the most efficient use of the information available in internal validation studies and calibration studies \cite{Carroll2006}. More efficient methods for measurement error correction methods are therefore implemented in \pkg{mecor}. A more efficient RC estimator, called \textit{efficient RC}, was introduced by Spiegelman et al. \cite{Spiegelman2001ERC}. A more efficient MM estimator was introduced by Keogh et al. \cite{Keogh2016}, which is called the Buonaccorsi approach using the method of moments. For simplicity, we will refer to this method as \textit{efficient MM}. 

Likewise, in replicates studies, \textit{standard RC} does not make the most efficient use of the information available \cite{Frost2000}. The \textit{standard RC} method is sub-optimal in terms of efficiency, since the method depends on the ordering of the replicate measurements \cite{Frost2000}. This can be intuitively understood as follows. The \textit{standard RC} regresses the mean of all but the first replicate on the first replicate, but this could as easily be exchanged with the second replicate. Therefore, different approaches are possible (e.g., maximum likelihood) \cite{Frost2000}. \cite{Bartlett2009LMM} showed how a standard random-intercepts model can be used to obtain \textit{maximum likelihood} (ML) estimates that are more efficient than \textit{standard RC}, at the cost of some additional parametric assumptions, discussed in section \ref{sec:meascor:mle}. 

Section \ref{sec:meascor:rc} introduces \textit{standard RC} and \textit{validation RC} for covariate measurement error correction, and \textit{standard MM} for outcome measurement error correction. \textit{Efficient RC} and \textit{efficient MM} are introduced in section \ref{sec:meascor:erc} and the maximum likelihood approach for replicates studies is introduced in section \ref{sec:meascor:mle}. When no information is available to estimate the parameters of the measurement error model, a \textit{sensitivity analysis} or \textit{quantitative bias analysis} can be used to analyse the sensitivity of study results to measurement error \cite{Lash2009, Nab2020}. An approach for conducting \textit{sensitivity analyses} is discussed in section \ref{sec:meascor:sensana}.

\subsection{Standard measurement error correction}\label{sec:meascor:rc}
\subsubsection{Covariate measurement error}
In \textit{standard RC}, the biased least squares estimator $\boldsymbol{\hat{\beta}^*}$ is multiplied by the inverse of an estimate of the calibration model matrix $\boldsymbol{{\Lambda}}$ to give a consistent and unbiased estimator of $\boldsymbol{\beta}$, denoted $\boldsymbol{\hat{\beta}}_{\text{RC}}$:
\begin{eqnarray}\label{eq:standardrc}
    \boldsymbol{\hat{\beta}}_{\mathrm{RC}} = \boldsymbol{\hat{\beta}^* \hat{\Lambda}}^{-1}
\end{eqnarray}
\textit{Standard RC} can be applied using all four types of validation studies (section \ref{sec:me:datastruc}).

To construct the calibration model matrix $\boldsymbol{\Lambda}$ (see equation (\ref{eq:calmodmatrix})), we estimate its components $\boldsymbol{\lambda}=(\lambda_{X^*}, \lambda_0, \boldsymbol{\lambda}_Z)$, from the linear calibration model:
\begin{eqnarray}\label{eq:calmod}
\E(X|X^*,\boldsymbol{Z}) = \lambda_{X^*} X^* + \lambda_0 + \boldsymbol{\lambda_Z} \boldsymbol{Z}',
\end{eqnarray} 
using least squares. Here, $\boldsymbol{\lambda_Z}$ is a $1 \times k$ matrix. Throughout, we assume that the calibration model matrix is correctly specified. To obtain estimates of the parameters of interest $\boldsymbol{\lambda}$ in an internal validation study (Table \ref{tab:covvalstudy}) and external validation study (Table \ref{tab:exvalstudy:cov}), the error-free reference measure $X$ is regressed on the error-prone substitute measure $X^*$ and the other covariates $\boldsymbol{Z}$. To obtain estimates of the parameters of interest $\boldsymbol{\lambda}$ in a replicates study (Table \ref{tab:repstudy:wide}), the mean of all replicates except the first replicate (i.e., $X^{*_r}_2, \dots, X^{*_r}_m$) is regressed on the first replicate $X^*_1$ and the other covariates $\boldsymbol{Z}$. To obtain estimates of the parameters of interest $\boldsymbol{\lambda}$ in a calibration study (Table \ref{tab:covcalstudy}), the mean of the replicates $X^{*_r}_1, \dots, X^{*_r}_m$ with random measurement error is regressed on the measurement $X^{*_s}$ with systematic measurement error and the other covariates $\boldsymbol{Z}$.

An adapted version of \textit{standard RC} in internal validation studies is \textit{validation RC} \cite{Carroll2006}. In \textit{validation RC}, the outcome $Y$ is regressed on the calibrated values $X_{\mathrm{cal}}$ and $\boldsymbol{Z}$. The calibrated values $X_{\mathrm{cal}}$ are constructed as follows: if $X$ is observed, $X_{\mathrm{cal}} = X$, and if $X$ is not observed, $X_{\mathrm{cal}} = \E(X|X^*,\boldsymbol{Z})$. The parameters from the regression of $Y$ on $X_{\mathrm{cal}}$ and $\boldsymbol{Z}$ are estimates of our parameters of interest $\boldsymbol{\beta}$ in equation (\ref{eq:outmem}). Note that \textit{standard RC} described above is identical to using $X_{\mathrm{cal}} = \E(X|X^*,\boldsymbol{Z})$ for all $X$ \cite{KeoghSTRATOS:2020}.

\subsubsection{Outcome measurement error}
In \textit{standard MM}, the biased least squares estimator $\boldsymbol{\hat{\beta}^*}$ is multiplied by the inverse of an estimate of the outcome measurement error model matrix $\boldsymbol{\Theta}$ to give a consistent and unbiased estimator of $\boldsymbol{\beta}$, denoted $\boldsymbol{\hat{\beta}}_{\text{MM}}$:
\begin{eqnarray}\label{eq:standardmm} 
\boldsymbol{\hat{\beta}}_{\text{MM}} = \boldsymbol{(\hat{\beta}^*,1) \hat{\Theta}}^{-1}.
\end{eqnarray}
\textit{Standard MM} can be applied using internal and external validation studies, and calibration studies (section \ref{sec:me:datastruc}).

To construct the outcome measurement error model matrix $\boldsymbol{\Theta}$ (see equation (\ref{eq:outmematrix})), we estimate its components $\boldsymbol{\theta} = (\theta_0,\theta_1)$ from the linear measurement error model $\E(Y^*|Y)=\theta_0 + \theta_1 Y$ using least squares. Throughout, we assume that the measurement error model matrix is correctly specified. To obtain estimates of the parameters of interest $\boldsymbol{\theta}$ in an internal validation study (Table \ref{tab:outvalstudy}) and an external validation study (Table \ref{tab:exvalstudy:out}), the error-prone substitute measurement $Y^*$ is regressed on the error-free reference measurement $Y$. To obtain estimates of the parameters of interest $\boldsymbol{\theta}$ in a calibration study (Table \ref{tab:outcalstudy}), the measurement $Y^{*_s}$ with systematic measurement error is regressed on the mean of the replicates $Y^{*_r}_1, \dots, Y^{*_r}_m$ with random measurement error.

\subsubsection{Differential outcome measurement error in univariable analyses} 
For the special case of differential measurement error, the outcome measurement error model matrix $\boldsymbol{\Theta}$ (see equation (\ref{eq:diffmematrix})), can be constructed as follows. We estimate its components $\boldsymbol{\theta}$ = ($\theta_{00}$, $\theta_{01}$, $\theta_{10}$, $\theta_{11}$) from the measurement error model $\E(Y^*|Y,X)=\theta_{00} + (\theta_{01} -\theta_{00})X + \theta_{10}Y + (\theta_{11}-\theta_{10})XY$. This model can be fitted directly in an internal validation study (Table \ref{tab:outvalstudy}), provided that the random internal subset includes exposed (i.e., $X=1$) and non-exposed individuals (i.e., $X=0$). The model can be fitted in an external validation study (Table \ref{tab:exvalstudy:out}), provided that $X$ is measured, and that exposed and non-exposed individuals are included in the external set. In a calibration study (Table \ref{tab:outcalstudy}), the measurement with systematic measurement error is regressed on the mean of the replicates $Y^{*_r}_1, \dots, Y^{*_r}_m$ with random measurement error and the covariate $X$ (again, provided that the random subset with replicates with random measurement error includes exposed and non-exposed individuals).

\subsubsection{Variance estimation}
The variance of the \textit{standard RC} estimator can be estimated using the multivariate delta method \cite{Rosner1990} or the zero-variance method \cite{Franz2007Ratios}. Confidence intervals can then be obtained by constructing Wald-type confidence intervals using one of the former two methods. Additionally, confidence intervals can be obtained by the stratified bootstrap, by sampling the observations in the internal subset separately from the observations outside the internal subset. The variance of the \textit{standard MM} estimator can also be estimated with the multivariate delta method, the zero-variance method or the stratified bootstrap. Additionally, for \textit{standard RC}, confidence intervals for $\hat{\beta_X}_{\text{RC}}$ (the first element of the $\boldsymbol{\hat{\beta}_{\text{RC}}}$) can be obtained by the Fieller method \cite{Frost2000}. For \textit{standard MM}, confidence intervals for $\hat{\beta_X}_{\text{MM}}$ and $\boldsymbol{\hat{\beta_Z}_{\text{MM}}}$ (the first two elements of the $\boldsymbol{\hat{\beta}_{\text{MM}}}$) can be obtained by the Fieller method \cite{Nab2018}. Details of these procedures can be found in the appendix section \ref{app:sec:variance_rc}.

\subsection{More efficient measurement error correction}\label{sec:meascor:erc}
\subsubsection{Covariate measurement error}
\textit{Efficient RC} can be used in internal validation studies or calibration studies \cite{Spiegelman2001ERC}. It pools the \textit{standard RC} estimate with an internal estimate for $\boldsymbol{\beta}$ obtained in the internal validation study or calibration study. 

In internal validation studies, the error-free reference covariate $X$ is obtained in an internal subset of the main study (Table \ref{tab:covvalstudy}). By regressing the outcome $Y$ on $X$ and the other covariates $\boldsymbol{Z}$ using least squares in the internal subset, one obtains an unbiased estimate for our parameters of interest $\boldsymbol{\beta}$. Denote this estimator by $\boldsymbol{\hat{\beta}}_{\text{I}}$. This internal estimator $\boldsymbol{\hat{\beta}}_{\text{I}}$ can then be combined with the standard RC estimator $\boldsymbol{\hat{\beta}}_{\text{RC}}$ defined in equation (\ref{eq:standardrc}), by taking the inverse variance weighted mean of the two estimates:
\begin{eqnarray}\label{eq:intvalstudy:erc}
\boldsymbol{\hat{\beta}}_{\text{ERC}}=[\boldsymbol{\hat{\Sigma}}_{\beta_{\text{RC}}}^{-1} + \boldsymbol{\hat{\Sigma}}_{\beta_{\text{I}}}^{-1}]^{-1}[\boldsymbol{\hat{\Sigma}}_{\beta_{\text{RC}}}^{-1}\hat{\beta_{\text{RC}}} + \boldsymbol{\hat{\Sigma}}_{\beta_{\text{I}}}^{-1}\hat{\beta}_{\text{I}}],
\end{eqnarray}
where $\boldsymbol{\hat{\Sigma}}_{\beta_{\text{RC}}}^{-1}$ is the variance--covariance matrix obtained from the multivariate delta method and $\boldsymbol{\hat{\Sigma}}_{\beta_{\text{I}}}$ is the standard variance--covariance matrix of  a least squares estimator. The efficient RC estimator defined above is an unbiased, consistent and the most efficient estimator for $\boldsymbol{\beta}$ if sampling into the internal validation set is unbiased (e.g., if the validation study is a random subset of participants) \cite{Spiegelman2001ERC}.

In calibration studies, the covariate $X$ is observed with random measurement error in an internal subset of the main study (Table \ref{tab:covcalstudy}). If at least 2 replicates are available, an unbiased estimator for $\boldsymbol{\beta}$ can be obtained by using the standard RC estimator for a replicates study (see section \ref{sec:meascor:rc}) in the internal subset. Again, denote this estimator by $\boldsymbol{\hat{\beta}}_{\text{I}}$. Then, the estimate obtained from the internal subset can be pooled with the standard RC estimate following equation (\ref{eq:intvalstudy:erc}). Alternatively, an unbiased estimator for $\boldsymbol{\beta}$ using the replicates in the internal subset can be obtained by using the ML estimation discussed in section \ref{sec:meascor:mle}. Again, this estimate can then be pooled with the standard RC estimate following equation (\ref{eq:intvalstudy:erc}).

\subsubsection{Outcome measurement error} 
\textit{Efficient MM} can be used in internal validation studies or calibration studies \cite{Keogh2016}. It pools the \textit{standard MM} estimate with an internal estimate for $\boldsymbol{\beta}$ obtained in the internal validation study or calibration study. 

In internal validation studies, the error-free reference outcome $Y$ is obtained in an internal subset of the main study (Table \ref{tab:outvalstudy}). By regressing $Y$ on the covariates $X$ and $\boldsymbol{Z}$ using least squares in the internal subset, one obtains an unbiased estimator for $\boldsymbol{\beta}$. Denote this estimator by $\boldsymbol{\hat{\beta}}_{\text{I}}$. In calibration studies, the outcome is observed with random measurement error in an internal subset of the main study (Table \ref{tab:outcalstudy}). The internal estimator $\boldsymbol{\hat{\beta}}_{\text{I}}$ is obtained by regressing the outcome $Y^{*,r}$ with random measurement error on the covariates $X$ and $\boldsymbol{Z}$ using least squares in the internal subset. Using the outcome with random measurement error will lead to the unbiased estimation of the association under study since random outcome measurement error does not bias the association. A single measurement with random measurement error (i.e., $m = 1$ in Table \ref{tab:outvalstudy}) is sufficient to obtain an internal estimate. However, if the outcome with random measurement error is observed more than once, the mean of the measures $Y^{*_r}_1, \dots, Y^{*_r}_m$ can be used and regressed on the covariates $X$ and $\boldsymbol{Z}$. Subsequently, the estimate obtained from the internal subset in an internal validation study or calibration study can be pooled with the \textit{standard MM} estimate following equation (\ref{eq:intvalstudy:erc}), by replacing the \textit{standard RC} estimate with the \textit{standard MM} estimate in the equation.

\subsubsection{Differential outcome measurement error in univariable analyses}
In internal validation studies, the internal estimator $\boldsymbol{\hat{\beta}}_{\text{I}}$ can be obtained by regressing $Y$ on the covariates $X$ and $\boldsymbol{Z}$ using least squares. In calibration studies, the internal estimator $\boldsymbol{\hat{\beta}}_{\text{I}}$ can be obtained by regressing the outcome $Y^{*,r}$ with random measurement error on the covariates $X$ and $\boldsymbol{Z}$. A single measurement with random measurement error (i.e., $m = 1$ in Table \ref{tab:outvalstudy}) is sufficient to obtain an internal estimate. However, if the outcome with random measurement error is observed more than once, the mean of the measures $Y^{*_r}_1, \dots, Y^{*_r}_m$ can be used and regressed on the covariates $X$ and $\boldsymbol{Z}$. We assume that the internal subset is a random subset of the main study, and hence that exposed and unexposed are included in the internal subset. Subsequently, the estimate obtained from the internal subset in an internal validation study or calibration study can be pooled with the \textit{standard MM} estimate following equation (\ref{eq:intvalstudy:erc}), by replacing the \textit{standard RC} estimate with the \textit{standard MM} estimate in the equation.

\subsubsection*{Variance estimation}
The variance of the \textit{efficient RC} estimator can be obtained from the following:
\begin{eqnarray*}
\boldsymbol{\hat{\Sigma}}_{\beta_{\text{ERC}}} = [\boldsymbol{\hat{\Sigma}}_{\beta}^{-1} + \boldsymbol{\hat{\Sigma}}_{\beta_{\text{I}}}^{-1}]^{-1}.
\end{eqnarray*}
The variance of the \textit{efficient RC} estimator can also be obtained by stratified bootstrapping, by sampling the observations in the internal subset separately from the observations outside the internal subset. Confidence intervals can be obtained by constructing Wald-type confidence intervals using one of the former two variances or by stratified percentile bootstrap. The same applies for the \textit{efficient MM} estimator. 

\subsection{Maximum likelihood estimation for replicates studies}\label{sec:meascor:mle}
The use of a standard random-intercepts model to obtain maximum likelihood (ML) estimates for $\boldsymbol{\beta}$ in replicates studies was introduced by \cite{Bartlett2009LMM}. To explain the \textit{ML} method for replicates studies, we add the index $i = 1, \dots, n$ to our notation in the outcome model:
\begin{equation*}
    Y_i = \beta_X X_i + \beta_0 + \boldsymbol{\beta}_Z \boldsymbol{Z}_i' + e_i, \qquad \mathrm{Var}(e_i)=\sigma^2,
\end{equation*}
where we again assume that $\E(e_i)=0$ and $\mathrm{Cov}(e_i, X_i) = \mathrm{Cov}(e_i, \boldsymbol{Z}_i) = 0$. Further, $\boldsymbol{Z}_i = (Z_{i1}, \dots, Z_{ik})$ and $\boldsymbol{\beta}_Z$ is again a $1 \times k$ matrix. On top of these assumptions, we also assume that the $e_i$ are normal and independently distributed. Additionally, assume that $X_i$ is normally distributed given $\boldsymbol{Z}_i$, with, 
\begin{eqnarray*}\label{eq:meanxgivenz}
    \E(X_i|\boldsymbol{Z}_i) = \rho_0 + \boldsymbol{\rho_Z}\boldsymbol{Z}_i' \quad \text{and} \quad \mathrm{Var}(X_i|\boldsymbol{Z}_i) = \sigma^2_{X_i|\boldsymbol{Z}_i},
\end{eqnarray*}
where $\boldsymbol{\rho}_Z$ is a $1 \times k$ matrix. In a replicates study, $X_i$ is not observed. Instead, $m$ replicates of the error-prone measurement $\boldsymbol{X}^{*_r}_{i}=(X^{*_r}_{i1}, \dots, X^{*_r}_{im})$ are observed, for $i = 1, \dots, n$. In a full-replicates study (Table \ref{tab:repstudy:wide}), we assume that the number of replicate measurements $m \geq 2$ is constant for every individual. In a partial-replicates study (Table \ref{tab:repstudy:long}), we assume that the number of replicates $m \geq 2 $ is constant in the replicate sub-study and $m=1$ in the main study. These measurements are assumed to follow the following random measurement error model:
\begin{equation*}
    X^{*_r}_{ij} = X_{i} + U_{ij}, \qquad \mathrm{Var}(U_{ij})=\tau^2, \qquad j=1, \dots, m,
\end{equation*}
where we again assume that $\E(U_{ij})=0$, $\mathrm{Cov}(U_{ij}, X_i)=0$, and that the measurement error in non-differential, i.e., the errors $U_{ij}$ are independent of the errors $e_i$ in the outcome model described above. In addition, we also assume that the errors $U_{ij}$ are normal and independently distributed.

We consider the likelihood function when only $Y_i$, $\boldsymbol{X}^{*_r}_i$ and $\boldsymbol{Z}_i$ are observed. The log likelihood can be factorized as follows:
\begin{eqnarray}\label{eq:loglikelihood}
    \ell(\boldsymbol{\theta}|Y_i, \boldsymbol{X}^{*_r}_i, \boldsymbol{Z}_i) = \log (f(Y_i|\boldsymbol{Z}_i,\boldsymbol{\theta})) + \log(f(\boldsymbol{X}^{*_r}_i|Y_i, \boldsymbol{Z}_i, \boldsymbol{\theta})),
\end{eqnarray}
where $\boldsymbol{\theta} = (\beta_X, \beta_0, \boldsymbol{\beta}_Z, \sigma^2, \rho_0, \boldsymbol{\rho_{Z}}, \sigma^2_{X|\boldsymbol{Z}}, \tau^2)$. From the assumptions that $X_i|\boldsymbol{Z_i}$ is normally distributed, the $e_i$ are normally distributed and that $X_i|\boldsymbol{Z_i}$ and $e_i$ are independent, \cite{Bartlett2009LMM} show that $Y_i$ given $\boldsymbol{Z}_i$ is normal with mean $\delta_0 + \boldsymbol{\delta_Z} \boldsymbol{Z}_i$ and variance $\sigma^2_{Y|\boldsymbol{Z}}$, where $\boldsymbol{\delta}_Z$ is a $1 \times k$ matrix. Furthermore, since $X_i|\boldsymbol{Z_i}$ and $Y_i|\boldsymbol{Z_i}$ are jointly normal, $X_i|Y_i,\boldsymbol{Z}_i$ is also normal. \cite{Bartlett2009LMM} show that we can therefore write:
\begin{equation*}
    X_i = \kappa_0 + \kappa_Y Y_i + \boldsymbol{\kappa_{Z}}\boldsymbol{Z}_i + b_i,
\end{equation*}
where $b_i \sim \mathrm{N}(0, \sigma^2_{X|Y,\boldsymbol{Z}})$. Then, since $X^*_{ij} = X_i + U_{ij}$, it follows from the above equation that,
\begin{equation*}
    X^*_{ij} = \kappa_0 + \kappa_Y Y_i + \boldsymbol{\kappa_{Z}}\boldsymbol{Z}_i + b_i + U_{ij},
\end{equation*}
where $U_{ij} \sim N(0, \tau^2)$ is independent of $b_i$ \cite{Bartlett2009LMM} and $\boldsymbol{\kappa}_Z$ is a $1 \times k$ matrix. Hence, $\boldsymbol{X}_i^{*_r}$ given $Y_i$ and $\boldsymbol{Z}_i$ follows a random-intercepts model with fixed effects of $Y_i$ and $\boldsymbol{Z}_i$, random intercepts variance $\sigma^2_{X|Y,\boldsymbol{Z}}$ and within subject variance $\tau^2$. 

The parameter vector $\boldsymbol{\zeta}=(\delta_0, \boldsymbol{\delta_Z}, \sigma^2_{Y|\boldsymbol{Z}}, \kappa_0, \kappa_Y, \boldsymbol{\kappa_Z}, \sigma^2_{X|Y,\boldsymbol{Z}}, \tau^2)$ is a one-to-one function of the original model parameter vector $\boldsymbol{\theta} = (\beta_X, \beta_0, \boldsymbol{\beta}_Z, \sigma^2, \rho_0, \boldsymbol{\rho_{Z}}, \sigma^2_{X|\boldsymbol{Z}}, \tau^2)$. Accordingly, \cite{Bartlett2009LMM} show that the ML estimate for $\zeta$ can be obtained by maximizing the two likelihood components of equation (\ref{eq:loglikelihood}) separately. The likelihood component corresponding to $f(Y_i|\boldsymbol{Z_i}, \boldsymbol{\zeta})$ in equation (\ref{eq:loglikelihood}) can be maximized by fitting the least squares regression of $Y_i$ on $\boldsymbol{Z}_i$. The likelihood component corresponding to $f(\boldsymbol{X}_i^{*_r}|Y_i,\boldsymbol{Z}_i, \boldsymbol{\zeta})$ in equation (\ref{eq:loglikelihood}) can be maximized by fitting a random-intercepts model for $\boldsymbol{X}^{*_r}_i$ given $Y_i$ and $\boldsymbol{Z}_i$.

An ML estimate for $\boldsymbol{\beta}$ can now be obtained by the following formulas:
\begin{eqnarray*}
\beta_X
&=& \kappa_Y \times \frac{\sigma^2_{Y|\boldsymbol{Z}}}{\sigma^2_{X|Y,\boldsymbol{Z}} + \kappa^2_Y\sigma^2_{Y|\boldsymbol{Z}}},\\
\beta_0 &=& \delta_0 - \beta_X\rho_0 = \delta_0 - \beta_X\{\kappa_0 + \kappa_Y\delta_0\},\\
\boldsymbol{\beta_Z} &=& \boldsymbol{\delta_Z}- \beta_X\boldsymbol{\rho_Z}=\boldsymbol{\delta_Z}- \beta_X \{\boldsymbol{\kappa_Z} + \kappa_Y\boldsymbol{\delta_Z} \}.
\end{eqnarray*}
The estimator $\boldsymbol{\hat{\beta}}_{\text{ML}} = (\hat{\beta}_{X_{\text{ML}}},\hat{\beta}_{0_{\text{ML}}}, \boldsymbol{\hat{\beta}}_{Z_{\text{ML}}})$ can be obtained by replacing the parameters from parameter vector $\boldsymbol{\zeta}$ by their estimates in the above equations. 

\subsubsection*{Variance estimation}
The variance of the maximum likelihood estimator can be estimated with the multivariate delta method \cite{Bartlett2009LMM}. Confidence intervals can then be obtained by constructing Wald-type confidence intervals. Confidence intervals can also be obtained by stratified bootstrap, by sampling the observations in the internal subset separately from the observations outside the internal subset. Details of these procedures can be found in the appendix section \ref{app:sec:variance_ml}.

\subsection{Sensitivity analyses}\label{sec:meascor:sensana}
Information from a validation study may not always be available. In that case, a formal correction is not possible. Nevertheless, when measurement error in a covariate or the outcome is expected, one may check how sensitive study results are to that measurement error. Literature or expert knowledge can be used to inform this sensitivity analysis, e.g., by hypothesizing possible ranges for the parameter values of the measurement model. 

When random covariate measurement error is expected, speculation is needed of the values of $\tau^2$, i.e., the variance of the random measurement error. Additionally, when systematic covariate measurement error is suspected, speculation is needed about the parameter values of the calibration model described by equation (\ref{eq:calmod}). When systematic outcome measurement error is suspected, speculation is needed about the parameter values of the outcome measurement error model, described in equation (\ref{eq:outmem}). 

\section[The R package mecor]{The \proglang{R} package \pkg{mecor}}
The \proglang{R} package \pkg{mecor} offers functionality to correct for measurement error in a continuous covariate or outcome in linear models with a continuous outcome. The main model fitting function in \pkg{mecor} is \code{mecor}:
\begin{Code}
mecor(formula, data, method, B)
\end{Code}
The function fits the linear model defined in \code{formula}, corrected for the measurement error in one of the variables. The arguments are as follows:
\begin{itemize}
    \item \code{formula} a \code{formula} object, with the response on the left of a `$\sim$' operator and the terms, separated by + operators, on the right. This argument takes the form \code{outcome ~ MeasError(substitute, reference, replicate, differential) + covariates} for covariate measurement error, and \code{MeasError(substitute, reference, replicate, differential) ~ covariates} for outcome measurement error. The \code{MeasError} object can be used for measurement error correction in internal validation, replicates and calibration studies. For external validation studies or sensitivity analyses of systematic measurement error, the object \code{MeasErrorExt(substitute, model)} is used instead of a \code{MeasError} object. For sensitivity analyses of random measurement error, the object \code{MeasErrorRandom(substitute, error)} is used.
    \item \code{method} specifies the method used for measurement error correction. The options are \code{"standard"} for standard RC and standard MM, \code{"valregcal"} for validation RC, \code{"efficient"} for efficient RC and efficient MM, and \code{"mle"} for maximum likelihood estimation.
    \item \code{B} number of bootstrap samples used for standard error estimation. The default is set to \code{0}.
\end{itemize}

An object of class \code{mecor} can be summarised using the \code{summary} function:
\begin{Code}
summary(object, alpha, zerovar, fieller)
\end{Code}
The arguments are as follows:
\begin{itemize}
    \item \code{object} an object of class \code{mecor}.
    \item \code{alpha} a numeric indicating the probability of obtaining a type II error. Defaults to \code{0.05}.
    \item \code{zerovar} a boolean indicating whether confidence intervals using the zero-variance method \cite{Franz2007Ratios} must be printed. Only available for \code{mecor} objects fitted with \code{method} equal to \code{"standard"}. Defaults to \code{FALSE}.
    \item \code{fieller} a boolean indicating whether confidence intervals using the fieller method \cite{Frost2000,Nab2018} must be printed. Only available for \code{mecor} objects fitted with \code{method} equal to \code{"standard"}. Defaults to \code{FALSE}.
\end{itemize}
The default \code{summary} object of an object of class \code{mecor} prints standard errors and confidence intervals obtained by the delta method. See the various `Variance estimation' paragraphs in section \ref{sec:meascor} for a description of the methods for variance estimation.

The \code{formula} argument in \code{mecor} contains a \code{MeasError} object, a \code{MeasErrorExt} object or a \code{MeasErrorRandom} object. All three objects are described below.

\subsection[The MeasError object]{The \code{MeasError} object}
To correct for measurement error using an internal validation study, a replicates study or a calibration study, the \code{formula} argument in \code{mecor} contains a \code{MeasError} object on the right-hand side (covariate measurement error) or left-hand side (outcome measurement error). The \code{MeasError} object can be used for random and systematic measurement error correction, depending on the method used to correct for the measurement error in \code{mecor}:
\begin{Code}
MeasError(substitute, reference, replicate, differential)
\end{Code}
with the arguments being described as follows:
\begin{itemize}
    \item \code{substitute} the error-prone substitute measurement; 
    \item \code{reference} the gold-standard reference measurement, to be used in case of an internal validation study, else \code{NULL};
    \item \code{replicate} (a vector of) the replicate measurement of the error-prone substitute measurement, to be used in case of a replicates study or calibration study, else \code{NULL};
    \item \code{differential} the binary exposure on which the outcome measurement error structure is dependent, to be used for differential outcome measurement error in univariable analyses, else \code{NULL}.
\end{itemize}
Depending on the type of validation study used, either argument \code{reference} (internal validation study) or \code{replicate} (replicates study or calibration study) should be used, but never both.

\subsection[The MeasErrorExt object]{The \code{MeasErrorExt} object}
To correct for measurement error using an external validation study, the \code{formula} object in \code{mecor} contains a \code{MeasErrorExt} object on the right-hand side (covariate measurement error) or left-hand side (outcome measurement error):
\begin{Code}
MeasErrorExt(substitute, model)
\end{Code}
with the arguments being described as follows:
\begin{itemize}
    \item \code{substitute} the error-prone measurement;
    \item \code{model} a fitted \code{lm} object of the calibration model in equation (\ref{eq:calmod}) (covariate measurement error) or the measurement error model in equation (\ref{eq:outmem}) (outcome measurement error). Or alternatively, a \code{list} with named arguments \code{coef} containing a vector of the coefficients of the calibration model or measurement error model and named argument \code{vcov} containing a matrix of the corresponding variance--covariance matrix. The argument \code{vcov} is not required.
\end{itemize}
The argument \code{model} is also used for conducting a sensitivity analysis by making informed guesses about the parameters of the calibration model (covariate measurement error) or measurement error model (outcome measurement error). 

\subsection[The MeasErrorRandom object]{The \code{MeasErrorRandom} object}
When random measurement error in a covariate is suspected but cannot be quantified, the \code{MeasErrorRandom} object can be used to conduct a sensitivity analysis:
\begin{Code}
MeasErrorRandom(substitute, variance)
\end{Code}
with the arguments being described as follows:
\begin{itemize}
    \item \code{substitute} the error-prone measurement;
    \item \code{variance} a numeric indicating the random measurement error variance in the substitute measurement, i.e., the parameter value of $\tau^2$ in equation (\ref{eq:covmem}).
\end{itemize}

\section{Examples}
Seven simulated datasets are included in the package \pkg{mecor}. These datasets represent the data structures described in section \ref{sec:me:datastruc}. There is an internal validation study with covariate measurement error (\code{icvs}), an internal validation study with outcome measurement error (\code{iovs}), an internal validation study with differential measurement error (\code{iovs\textunderscore diff}), a replicates study (\code{rs}) and a calibration study with covariate measurement error (\code{ccs}). The dataset \code{ecvs} provides an external validation study for the \code{icvs} dataset, and the dataset \code{eovs} provides an external validation study for the \code{iovs} dataset. These datasets are described and analysed in the following sections. 

\subsection{Internal validation study}
The dataset \code{icvs} is a simulated dataset, representing the structure of the internal covariate-validation study shown in Table \ref{tab:covvalstudy}. The dataset contains 1,000 observations of the continuous outcome \code{Y}, an error-prone continuous exposure \code{X\textunderscore star} and a continuous covariate \code{Z}. The error-free  measurement of the exposure \code{X} is available in approximately 25\% of the study population. 
\begin{CodeChunk}
\begin{CodeInput}
R> data("icvs", package = "mecor")
R> head(icvs)
\end{CodeInput}
\begin{CodeOutput}
          Y     X_star          Z          X
1 -3.473164 -0.2287010 -1.5858049         NA
2 -3.327934 -1.3320494 -0.6077454         NA
3  1.314735  2.0305727  0.4461727  2.2256377
4  1.328727  0.3027101  0.1739813         NA
5  1.240446 -0.8465389  1.5480392 -0.7521792
6  3.183868  0.1081888  1.1230232         NA
\end{CodeOutput}
\end{CodeChunk}
By ignoring the measurement error in \code{X\textunderscore star}, a linear model can be fitted to the data as follows:
\begin{CodeChunk}
\begin{CodeInput}
R> lm(Y ~ X_star + Z, data = icvs)
\end{CodeInput}
\begin{CodeOutput}
Call:
lm(formula = Y ~ X_star + Z, data = icvs)

Coefficients:
(Intercept)       X_star            Z  
   -0.03947      0.41372      2.08457 
\end{CodeOutput}
\end{CodeChunk}
The coefficients of this model will however be biased due to the measurement error in \code{X\textunderscore star}. The measurement error in \code{X\textunderscore star} can be corrected for using standard RC as follows: 
\begin{CodeChunk}
\begin{CodeInput}
R> mecor(Y ~ MeasError(X_star, reference = X) + Z, 
+        data = icvs, 
+        method = "standard")
\end{CodeInput}
\begin{CodeOutput}
Call:
mecor(formula = Y ~ MeasError(X_star, reference = X) + Z, data = icvs, 
method = "standard")

Coefficients Corrected Model:
(Intercept)           X           Z 
-0.03730093  0.49290832  2.00122724 

Coefficients Uncorrected Model:
(Intercept)      X_star           Z 
-0.03946702  0.41371614  2.08457045 
\end{CodeOutput}
\end{CodeChunk}
Stratified percentile bootstrap confidence intervals of the coefficients of the corrected model can be obtained by using the argument \code{B} in the function \code{mecor}. To obtain standard errors and confidence intervals using the Fieller method or zero-variance method, the arguments \code{zerovar} and \code{fieller} of the \code{summary} object are set to \code{TRUE}:
\begin{CodeChunk}
\begin{CodeInput}
R> set.seed(20210120)
R> mecor_fit <- mecor(Y ~ MeasError(X_star, reference = X) + Z, 
+                     data = icvs, 
+                     method = "standard", 
+                     B = 999)
R> summary(mecor_fit, zerovar = TRUE, fieller = TRUE)
\end{CodeInput}
\begin{CodeOutput}
Call:
mecor(formula = Y ~ MeasError(X_star, reference = X) + Z, data = icvs, 
    method = "standard", B = 999)

Coefficients Corrected Model:
             Estimate       SE SE (btstr) SE (zerovar)
(Intercept) -0.037301 0.035943   0.033803     0.033365
X            0.492908 0.036959   0.039326     0.034412
Z            2.001227 0.051999   0.054882     0.048431

95% Confidence Intervals:
             Estimate       LCI      UCI LCI (btstr) UCI (btstr)
(Intercept) -0.037301 -0.107748 0.033147   -0.099337    0.032160
X            0.492908  0.420470 0.565347    0.418986    0.573478
Z            2.001227  1.899311 2.103143    1.888301    2.108332
            LCI (zerovar) UCI (zerovar) LCI (fieller) UCI (fieller)
(Intercept)     -0.102695      0.028093            NA            NA
X                0.425463      0.560354      0.421772      0.566887
Z                1.906303      2.096151            NA            NA
Bootstrap Confidence Intervals are based on 999 bootstrap replicates using percentiles 

The measurement error is corrected for by application of regression calibration 

Coefficients Uncorrected Model:
             Estimate Std. Error t value Pr(>|t|)
(Intercept) -0.039467   0.033355 -1.1832    0.237
X_star       0.413716   0.028883 14.3239   <2e-16
Z            2.084570   0.044323 47.0316   <2e-16

95% Confidence Intervals:
             Estimate       LCI      UCI
(Intercept) -0.039467 -0.104921 0.025987
X_star       0.413716  0.357038 0.470394
Z            2.084570  1.997594 2.171547

Residual standard error: 1.052587 on 997 degrees of freedom
\end{CodeOutput}
\end{CodeChunk}
In addition to standard RC, efficient RC (\code{method = "efficient"}) or validation RC (\code{method = "valregcal"}) can also be used to correct for the measurement error in the error-prone covariate \code{X\textunderscore star}.

The dataset \code{iovs} is a simulated dataset, representing the structure of the internal outcome-validation study shown in Table \ref{tab:outvalstudy}. The dataset contains 1,000 observations of the error-prone continuous outcome \code{Y\textunderscore star}, an continuous exposure \code{X} and a continuous covariate \code{Z}. The error-free  measurement of the outcome \code{Y} is taken in approximately 25\% of the study population.
\begin{CodeChunk}
\begin{CodeInput}
R> data("iovs", package = "mecor")
R> head(iovs)
\end{CodeInput}
\begin{CodeOutput}
       Y_star          X            Z          Y
1 -1.47206289 -1.1601736 -1.031101614 -2.5259806
2 -1.17635031 -0.4880010 -0.411765610         NA
3 -0.04871530  0.4211575  0.002352782  0.9289676
4  0.31207120 -1.2179780  0.029320987         NA
5 -0.05210891 -0.6253944  0.596399480         NA
6 -2.31198582 -2.2696848 -1.174442382         NA
\end{CodeOutput}
\end{CodeChunk}
The measurement error in \code{Y\textunderscore star} can be accounted for by using standard MM as shown in the following:
\begin{CodeChunk}
\begin{CodeInput}
R> mecor(MeasError(Y_star, reference = Y) ~ X + Z, 
+        data = iovs, 
+        method = "standard")
\end{CodeInput}
\begin{CodeOutput}
Call:
mecor(formula = MeasError(Y_star, reference = Y) ~ X + Z, data = iovs, 
method = "standard")

Coefficients Corrected Model:
 (Intercept)            X            Z 
-0.009033377  0.499965568  2.012418865 

Coefficients Uncorrected Model:
(Intercept)           X           Z 
0.005138783 0.244062431 0.982379329 
\end{CodeOutput}
\end{CodeChunk}
In addition to standard MM, efficient MM (\code{method = "efficient"}) can also be used to correct for the measurement error in the error-prone outcome \code{Y\textunderscore star}.

The dataset \code{iovs\textunderscore diff} is a simulated dataset, again representing the structure of an internal outcome-validation study. The dataset contains 1,000 observations of the error-prone continuous outcome \code{Y\textunderscore star} with differential measurement error and a binary exposure \code{X}. The error-free  measurement of the outcome \code{Y} is taken in approximately 25\% of the study population.
\begin{CodeChunk}
\begin{CodeInput}
R> data("iovs_diff", package = "mecor")
R> head(iovs_diff)
\end{CodeInput}
\begin{CodeOutput}
       Y_star X          Y
1 -0.67852444 0 -1.1510182
2  1.58800417 1  0.8830208
3  0.13590696 0         NA
4  1.26666798 1         NA
5 -0.07451448 0         NA
6  2.25628302 1  0.6931214
\end{CodeOutput}
\end{CodeChunk}
The differential measurement error in \code{Y\textunderscore star} can be accounted for by using the following:
\begin{CodeChunk}
\begin{CodeInput}
R> mecor(MeasError(Y_star, reference = Y, differential = X) ~ X, 
+        data = iovs_diff, 
+        method = "standard")
\end{CodeInput}
\begin{CodeOutput}
Call:
mecor(formula = MeasError(Y_star, reference = Y, differential = X) ~ 
    X, data = iovs_diff, method = "standard")

Coefficients Corrected Model:
(Intercept)           X 
-0.01673621  0.51033737 

Coefficients Uncorrected Model:
 (Intercept)            X 
-0.005200984  1.738188190 
\end{CodeOutput}
\end{CodeChunk}
Efficient MM (\code{method = "efficient"}) can also be used to correct for the differential measurement error in the error-prone outcome \code{Y\textunderscore star}.

\subsection{Replicates study}
The dataset \code{rs} is a simulated dataset, representing the structure of a replicates study, shown in Table \ref{tab:repstudy:wide}. The dataset contains 1,000 observations of the continuous outcome \code{Y}, three replicate measures of a continuous error-prone exposure \code{X\textunderscore star\textunderscore 1}, \code{X\textunderscore star\textunderscore 2} and \code{X\textunderscore star\textunderscore 3}, and covariates \code{Z1} and \code{Z2}. 
\begin{CodeChunk}
\begin{CodeInput}
R> data("rs", package = "mecor")
R> head(rs)
\end{CodeInput}
\begin{CodeOutput}
           Y  X_star_1  X_star_2   X_star_3         Z1         Z2
1  4.3499125 2.3579436  2.109616  2.5476139  1.3052590  1.0955135
2  1.5012541 2.4388260  2.951810  1.1754237  0.5207398  0.1326150
3  0.8249127 2.1473954  1.640624  1.1649830 -0.2035013 -0.2137111
4  5.5258514 2.8678562  2.547374  3.7509709  2.2994651  0.3441749
5  1.9177871 1.1964137  2.280420  2.1699911 -0.2930064  1.9122899
6 -1.0732459 0.8809592 -1.246339 -0.0832355 -0.8107538  0.7414892
\end{CodeOutput}
\end{CodeChunk}
The measurement error in the error-prone exposure \code{X\textunderscore star \textunderscore 1} can be accounted for as follows:
\begin{CodeChunk}
\begin{CodeInput}
R> mecor(Y ~ MeasError(X_star_1, replicate = cbind(X_star_2, X_star_3) + Z1 + Z2),
+     data = rs, 
+     method = "standard")
\end{CodeInput}
\begin{CodeOutput}
Call:
mecor(formula = Y ~ MeasError(X_star_1, replicate = cbind(X_star_2, 
    X_star_3)) + Z1 + Z2, data = rs, method = "standard")

Coefficients Corrected Model:
 (Intercept) cor_X_star_1           Z1           Z2 
 -0.07416626   0.54994384   1.95079727   0.98285575 

Coefficients Uncorrected Model:
(Intercept)    X_star_1          Z1          Z2 
-0.03821319  0.44013564  2.04396587  1.08129753 
\end{CodeOutput}
\end{CodeChunk}
Maximum likelihood estimation (\code{method = "mle"}) can also be used to correct for the measurement error in the error-prone exposure \code{X1\textunderscore star}. Note that, in this example dataset, the coefficients of the corrected model using standard RC will differ when \code{MeasError(X2\textunderscore star, replicate = cbind(X1\textunderscore star, X3\textunderscore star)} is used instead of \code{MeasError(X1\textunderscore star, replicate = cbind(X2\textunderscore star, X3\textunderscore star)}. In contrast, the corrected estimated coefficients obtained using maximum likelihood estimation will not change when the order of replicates is changed. 

\subsection{Calibration study}
The dataset \code{ccs} is a simulated dataset, representing the structure of the covariate calibration study, shown in Table \ref{tab:outcalstudy}. The dataset contains 1,000 observations of the continuous outcome \code{Y}, an error-prone continuous exposure \code{X\textunderscore star} measured with systematic measurement error, two replicate measures of the same error-prone exposure \code{X\textunderscore star\textunderscore 1} and \code{X\textunderscore star\textunderscore 2} with random measurement error, and a covariate \code{Z}. The replicate measures \code{X\textunderscore star\textunderscore 1} and \code{X\textunderscore star\textunderscore 2} are only observed in the first 500 study participants.
\begin{CodeChunk}
\begin{CodeInput}
R> data("ccs", package = "mecor")
R> ccs[c(1, 500, 501, 1000), ]
\end{CodeInput}
\begin{CodeOutput}
              Y    X_star          Z  X_star_1   X_star_2
1    -0.3406228 -1.538746 -0.0936835 0.0314241 -1.1768822
500  -0.7107400  1.217544 -0.5028586 1.1228869  0.9426396
501   1.7714124  3.758594  0.1005625        NA         NA
1000  3.0392145  3.793622  1.1303447        NA         NA
\end{CodeOutput}
\end{CodeChunk}
The measurement error in the error-prone exposure \code{X\textunderscore star} can be accounted for as follows:
\begin{CodeChunk}
\begin{CodeInput}
R> mecor(Y ~ MeasError(X_star, replicate = cbind(X_star_1, X_star_2)) + Z, 
+        data = ccs,
+        method = "standard")
\end{CodeInput}
\begin{CodeOutput}
Call:
mecor(formula = Y ~ MeasError(X_star, replicate = cbind(X_star_1, 
    X_star_2)) + Z, data = ccs, method = "standard")

Coefficients Corrected Model:
(Intercept)  cor_X_star           Z 
 0.03148803  0.47565243  2.10394044 

Coefficients Uncorrected Model:
(Intercept)      X_star           Z 
 0.02212343  0.22725887  2.10577516 
\end{CodeOutput}
\end{CodeChunk}
By only using the first 500 observations, maximum likelihood estimation can be used to correct for the measurement error, using only the replicate measures as follows:
\begin{CodeChunk}
\begin{CodeInput}
R> mecor(Y ~ MeasError(X1_star, replicate = X2_star) + Z, 
+        data = ccs[1:500,], 
+        method = "mle")
\end{CodeInput}
\begin{CodeOutput}
Call:
mecor(formula = Y ~ MeasError(X1_star, replicate = X2_star) + 
    Z, data = ccs[1:500, ], method = "mle")

Coefficients Corrected Model:
(Intercept) cor_X1_star           Z 
-0.06286467  0.51201688  2.05333600 

Coefficients Uncorrected Model:
(Intercept)     X1_star           Z 
-0.06288496  0.43184108  2.12827829 
\end{CodeOutput}
\end{CodeChunk}
A pooled estimate of the two estimates obtained by maximum likelihood estimation and standard regression calibration can be obtained by using \code{method = "efficient"}.

\subsection{External validation study}
The dataset \code{ecvs} is a simulated dataset, representing the structure of the external part of the external covariate-validation study shown in Table \ref{tab:exvalstudy:cov}. The dataset contains 100 observations of the error-free continuous exposure \code{X}, the error-prone exposure \code{X\textunderscore star} and a covariate \code{Z}.
\begin{CodeChunk}
\begin{CodeInput}
R> data("ecvs", package = "mecor")
R> head(ecvs)
\end{CodeInput}
\begin{CodeOutput}
           X     X_star           Z
1 -1.8177022 -1.0599121 -0.90610225
2 -0.6256011 -1.1090224 -2.02664356
3  1.9752408  1.3199730  1.37740119
4  0.3140878 -0.1107001  0.24907466
5  1.3191880  1.6079029 -0.52663051
6 -0.6149009 -0.7632676  0.03778671
\end{CodeOutput}
\end{CodeChunk}
Suppose that in the dataset \code{icvs}, the gold-standard measure \code{X} had not been observed. Using dataset \code{ecvs}, we can correct for the measurement error in \code{X\textunderscore star} in \code{icvs}. The first step is to fit the calibration model in the external validation study as follows: 
\begin{CodeChunk}
\begin{CodeInput}
R> calmod_fit <- lm(X ~ X_star + Z, data = ecvs)
R> calmod_fit
\end{CodeInput}
\begin{CodeOutput}
Call:
lm(formula = X ~ X_star + Z, data = ecvs)

Coefficients:
(Intercept)       X_star            Z  
   -0.05276      0.86693      0.18843 
\end{CodeOutput}
\end{CodeChunk}
The second step is to use the calibration model \code{calmod\textunderscore fit} in the \code{MeasErrorExt} object as follows:
\begin{CodeChunk}
\begin{CodeInput}
R> data("icvs", package = "mecor")
R> mecor(Y ~ MeasErrorExt(X_star, calmod_fit) + Z, 
+        data = icvs, 
+        method = "standard")
\end{CodeInput}
\begin{CodeOutput}
Call:
mecor(formula = Y ~ MeasErrorExt(X_star, calmod_fit) + Z, data = icvs, 
    method = "standard")

Coefficients Corrected Model:
(Intercept)  cor_X_star           Z 
-0.01428791  0.47721827  1.99464677 

Coefficients Uncorrected Model:
(Intercept)      X_star           Z 
-0.03946702  0.41371614  2.08457045 
\end{CodeOutput}
\end{CodeChunk}
Dataset \code{eovs} is a simulated dataset, representing the structure of the external part of the external outcome-validation study shown in Table \ref{tab:exvalstudy:cov}. The dataset contains 100 observations of the error-free  continuous outcome \code{Y} and the error-prone outcome \code{Y\textunderscore star}. 
\begin{CodeChunk}
\begin{CodeInput}
R> data("eovs", package = "mecor")
R> head(eovs)
\end{CodeInput}
\begin{CodeOutput}
           Y      Y_star
1 -0.7649570 -1.26057465
2 -3.2525931 -1.41926038
3  0.9778762 -0.08929029
4 -1.6535616 -0.29122347
5 -0.2961092 -0.21852096
6  2.9317563  2.13262253
\end{CodeOutput}
\end{CodeChunk}
Suppose that in the dataset \code{iovs}, the gold-standard measure of the outcome $Y$ was not observed. Using dataset \code{eovs}, we correct for the measurement error in \code{Y\textunderscore star} in \code{iovs}, by fitting the measurement error model, as follows:
\begin{CodeChunk}
\begin{CodeInput}
R> memod_fit <- lm(Y_star ~ Y, data = eovs)
R> data("iovs", package = "mecor")
R> mecor(MeasErrorExt(Y_star, memod_fit) ~ X + Z, 
+        data = iovs, 
+        method = "standard")
\end{CodeInput}
\begin{CodeOutput}
Call:
mecor(formula = MeasErrorExt(Y_star, memod_fit) ~ X + Z, data = iovs, 
    method = "standard")

Coefficients Corrected Model:
(Intercept)           X           Z 
0.008702382 0.473243397 1.904859053 

Coefficients Uncorrected Model:
(Intercept)           X           Z 
0.005138783 0.244062431 0.982379329 
\end{CodeOutput}
\end{CodeChunk}

\subsubsection{Sensitivity analyses}
Suppose that there is no error-free  measure and no external validation study available for dataset \code{icvs}. To investigate the sensitivity of study results to measurement error in variable \code{X\textunderscore star}, informed guesses of the coefficients of the calibration model are needed. Suppose one assumes that $E(X|X^*, Z)= 0.9X^* + 0.2Z$. A sensitivity analysis can then be conducted as follows:
\begin{CodeChunk}
\begin{CodeInput}
R> data("icvs", package = "mecor")
R> mecor_fit_sens <- mecor(Y ~ MeasErrorExt(X_star, list(coef = c(0, 0.9, 0.2))) + Z,
+                          data = icvs, 
+                          method = "standard")
R> mecor_fit_sens
\end{CodeInput}
\begin{CodeOutput}
Call:
mecor(formula = Y ~ MeasErrorExt(X_star, list(coef = c(0, 0.9, 
    0.2))) + Z, data = icvs, method = "standard")

Coefficients Corrected Model:
(Intercept)  cor_X_star           Z 
-0.03946702  0.45968460  1.99263353 

Coefficients Uncorrected Model:
(Intercept)      X_star           Z 
-0.03946702  0.41371614  2.08457045
\end{CodeOutput}
\end{CodeChunk}
The calibration model matrix used to correct for the measurement error in \code{X\textunderscore star}, is saved as \code{matrix} in the \code{corfit} object attached to \code{mecor\textunderscore fit\textunderscore sens}:
\begin{CodeChunk}
\begin{CodeInput}
R> mecor_fit_sens$corfit$matrix
\end{CodeInput}
\begin{CodeOutput}
        Lambda1 Lambda0 Lambda3
Lambda1     0.9       0     0.2
Lambda0     0.0       1     0.0
Lambda3     0.0       0     1.0
\end{CodeOutput}
\end{CodeChunk}
When random measurement error is expected in \code{X\textunderscore star}, a \code{MeasErrorRandom} object can be used, here in combination with zerovariance variance estimation of standard errors (assuming that there is no uncertainty in the speculated value of the variance of the random measurement error in \code{X\textunderscore star}):
\begin{CodeChunk}
\begin{CodeInput}
R > mecor_fit_random <- mecor(Y ~ MeasErrorRandom(X_star, variance = 0.25) + Z,
+                             data = icvs,
+                             method = "standard")
R > summary(mecor_fit_random, zerovar = T)
\end{CodeInput}
\begin{CodeOutput}
Call:
mecor(formula = Y ~ MeasErrorRandom(X_star, error = 0.25) + Z, 
    data = icvs, method = "standard")

Coefficients Corrected Model:
             Estimate SE (zerovar)
(Intercept) -0.032447     0.033390
cor_X_star   0.509533     0.035572
Z            1.985579     0.049241

95% Confidence Intervals:
             Estimate LCI (zerovar) UCI (zerovar)
(Intercept) -0.032447     -0.097890      0.032995
cor_X_star   0.509533      0.439813      0.579253
Z            1.985579      1.889067      2.082090

Coefficients Uncorrected Model:
             Estimate Std. Error t value Pr(>|t|)
(Intercept) -0.039467   0.033355 -1.1832    0.237
X_star       0.413716   0.028883 14.3239   <2e-16
Z            2.084570   0.044323 47.0316   <2e-16

95% Confidence Intervals:
             Estimate       LCI      UCI
(Intercept) -0.039467 -0.104921 0.025987
X_star       0.413716  0.357038 0.470394
Z            2.084570  1.997594 2.171547

Residual standard error: 1.052587 on 997 degrees of freedom
\end{CodeOutput}
\end{CodeChunk}
The calibration model matrix used to correct for the measurement error in \code{X\textunderscore star}, is again saved as \code{matrix} in the \code{corfit} object attached to \code{mecor\textunderscore fit\textunderscore sens}:
\begin{CodeChunk}
\begin{CodeInput}
R > mecor_fit_random$corfit$matrix
\end{CodeInput}
\begin{CodeOutput}
          Lambda1     Lambda2   Lambda3
Lambda1 0.8119518 -0.01377731 0.1942796
Lambda2 0.0000000  1.00000000 0.0000000
Lambda3 0.0000000  0.00000000 1.0000000
\end{CodeOutput}
\end{CodeChunk}
The sensitivity analyses can be expanded to ranges of possible coefficients of the calibration model or assumed variance of the random measurement error. 

\section{Conclusion}\label{sec:conclusion}
We demonstrated how measurement error correction methods can be applied using our \proglang{R} package \pkg{mecor}. These correction methods can be used in linear models with a continuous outcome when there is measurement error in the outcome or in a continuous covariate. The package accommodates measurement error correction methodology for a wide range of data structures: internal and external validation studies, replicates studies, and calibration studies. Various measurement error correction methods are implemented in the package: standard RC, efficient RC, validation RC and correction based on maximum likelihood estimation. For standard error estimation, the delta method and bootstrap are implemented for all methods. The package also facilitates sensitivity analysis or quantitative bias analysis when no data are available to estimate the parameters of the measurement error model, but the assumption of no measurement error is not warranted. We focused on studies in which interest lies in estimating a covariate-outcome association. In other types of studies, e.g., prediction studies, considerations for measurement error correction are different and may not even require corrections \cite{Luijken:2019, Luijken2019jce}. In future updates of the package, the measurement error correction methods may be extended to time-to-event \cite{Prentice1982} and binary outcomes, and multiple variables with measurement error \cite{Rosner1990,Rosner1992}. 

\section*{Computational details}

The results in this paper were obtained using
\proglang{R}~4.0.2. \proglang{R} itself
and \pkg{mecor} are available from the Comprehensive
\proglang{R} Archive Network (CRAN) at
\URL{https://CRAN.R-project.org/}.

\section*{Acknowledgments}
LN, MvS and RHHG were supported by grants from the Netherlands Organization for Scientific Research (ZonMW-Vidi project 917.16.430) and Leiden University Medical Center, LN was supported by Stichting Jo Kolk Studiefonds and Leids Universiteits Fonds in the form of a travel grant, RHK was supported by a Medical Research Council Methodology Fellowship (MR/M014827/1) and a UK Research and Innovation Future Leaders Fellowship (MR/S017968/1).

\bibliographystyle{unsrt}
\bibliography{refs.bib}


\newpage

\begin{appendix}
\section{Variance estimation}
\subsection{Standard regression calibration}\label{app:sec:variance_rc}
\textbf{Covariate measurement error.} The variance--covariance matrix for the standard regression estimator $\boldsymbol{\hat{\beta}_{\text{RC}}}$ can be approximated by using the multivariate delta method by \cite{Rosner1990}, given by
\begin{eqnarray}\label{app:eq:varcovrc_covme}
\boldsymbol{\hat{\Sigma}}_{\boldsymbol{\beta_{\text{RC}}}}(j_1,j_2) = (\boldsymbol{\hat{A}}'\boldsymbol{\hat{\Sigma}}_{\beta^*}\boldsymbol{\hat{A}})_{j_1,j_2} + \boldsymbol{\hat{\beta}^*}\boldsymbol{\hat{\Sigma}_{A,j_1,j_2}}\boldsymbol{\hat{\beta}^*}', \qquad j_1,j_2 = 1,\dots,(k+2),
\end{eqnarray}
where $\boldsymbol{\hat{A}}$ is the inverse of the calibration model matrix $\boldsymbol{\hat{\Lambda}}$. Further, $\boldsymbol{\hat{\Sigma}}_{\beta^*}$ is the variance--covariance matrix obtained from the naive regression defined in equation (\ref{eq:covmem}) in the main text and $\boldsymbol{\hat{\Sigma}_{A,j_1,j_2}}$ is the $(k+2)\times (k+2)$ matrix relating the $j_1$th and $j_2$th column of $\boldsymbol{A}$ (we refer to Appendix of \cite{Rosner1990} for a derivation). Additionally, the so-called zero-variance variance--covariance matrix for $\boldsymbol{\hat{\beta}}$ can be estimated by $\boldsymbol{\hat{A}}'\boldsymbol{\Sigma}_{\beta^*}\boldsymbol{\hat{A}}$ (i.e., by omitting the variance in the calibration model matrix).

A $100(1-\alpha)$ percent confidence interval for the $j$th element of $\boldsymbol{\hat{\beta}_{\text{RC}}}$ is then
\begin{eqnarray}\label{app:eq:waldint}
\boldsymbol{\hat{\beta}_{\text{RC}}}_j \pm \sqrt{\VAR(\boldsymbol{\hat{\beta}_{\text{RC}}}_j)},
\end{eqnarray}
where $\VAR(\boldsymbol{\hat{\beta}_{\text{RC}}}_j)$ is the jth element on the diagonal of $\boldsymbol{\hat{\Sigma}}_{\boldsymbol{\beta_{\text{RC}}}}$. The variance--covariance matrix $\boldsymbol{\hat{\Sigma}}_{\boldsymbol{\beta_{\text{RC}}}}$ can be obtained by either using the delta variance--covariance matrix or zero-variance variance--covariance matrix. In general, the zero-variance variance--covariance matrix will underestimate the true variance--covariance matrix and thus lead to too narrow confidence intervals.

Other methods to construct confidence intervals include stratified bootstrap \cite{Carroll2006} and the Fieller method \cite{Buonaccorsi2010, Franz2007Ratios, Frost2000, Nab2018}. In case of covariate measurement error, the Fieller method can only be applied to construct a $100(1-\alpha)$ percent confidence interval for the first element of $\boldsymbol{\hat{\beta}_{\text{RC}}}$, i.e., $\hat{\phi}_{\text{RC}}$. From \cite{Nab2018} we obtain:
\begin{eqnarray}\label{app:eq:fieller}
\{f_1 \pm \sqrt{f_1^2 - f_0f_2}/f_2\},
\end{eqnarray}
where $f_0=z_{\alpha/2}^2\VAR(\hat{\phi}^*)-\hat{\phi}^*$, $f_1= z_{\alpha/2}^2\COV(\hat{\phi}^*, \hat{\lambda}_1)- \hat{\phi}^* \hat{\lambda}_1$, $f_2=z_{\alpha/2}^2\VAR(\hat{\lambda}_1) - \hat{\lambda}_1^2$. Where it is assumed that $\COV(\hat{\phi}^*, \hat{\lambda}_1)$ is null. If the $(1-\alpha)\times 100 \%$ confidence interval of $\hat{\lambda}_1$ includes 0, the Fieller method does not lead to bounded confidence intervals. 
Bootstrap confidence intervals are obtained by sampling the people in the validation set separately from the people not included in the validation set \cite{Carroll2006} and taking the $(100-\alpha)$ percentiles of the obtained distribution. \\
\textbf{Outcome measurement error.} The variance--covariance matrix for the standard regression estimator $(\boldsymbol{\hat{\beta_{\text{RC}}}},1)$ can be approximated by applying the multivariate delta method similar to the variance obtained for the corrected estimator for covariate maesurement error,
\begin{eqnarray*}
\boldsymbol{\hat{\Sigma}}_{(\boldsymbol{\beta_{\text{RC}}},1)}(j_1,j_2) = (\boldsymbol{B}'\boldsymbol{\hat{\Sigma}}_{(\beta^*,1)}\boldsymbol{B})_{j_1,j_2} + (\boldsymbol{\hat{\beta}^*},1)\boldsymbol{\hat{\Sigma}_{B,j_1,j_2}}(\boldsymbol{\hat{\beta}^*},1)', \qquad j_1,j_2 = 1,\dots,(k+3),
\end{eqnarray*}
where $\boldsymbol{\hat{B}}$ is the inverse of the measurement error model matrix $\boldsymbol{\hat{\Theta}}$. $\boldsymbol{\hat{\Sigma}}_{(\beta^*,1)}$ is a $(k+3)\times(k+3)$ matrix where the upper $(k+2)\times(k+2)$ comprises the variance--covariance matrix obtained from the uncorrected regression defined by model (\ref{eq:genmodmeout}) and the last row and column contain zeros. Further, $\boldsymbol{\hat{\Sigma}_{B,j_1,j_2}}$ is the $(k+3)\times(k+3)$ matrix relating the $j_1$th and $j_2$th column of $\boldsymbol{B}$ (similar to \cite{Rosner1990}). The so-called zero-variance variance--covariance matrix for $\hat{\beta}$ can be estimated by $\boldsymbol{B}'\boldsymbol{\hat{\Sigma}}_{(\beta^*,1)}\boldsymbol{B}$. 

A $100(1-\alpha)$ percent confidence interval can be obtained from equation (\ref{app:eq:waldint}). Further, $100(1-\alpha)$ percent confidence intervals for $\hat{\phi}$ and $\boldsymbol{\hat{\gamma}}$ can be approximated by the Fieller method as defined in model \ref{app:eq:fieller}, where $f_0=\hat{\phi}^*-z_{\alpha/2}^2\VAR(\hat{\phi}^*)$, $f_1=\hat{\phi}^*/\hat{\theta}_1-z_{\alpha/2}^2\COV(\hat{\phi}^*, 1/\hat{\theta}_1)$, $f_2=1/\hat{\lambda}_1^2-z_{\alpha/2}^2\VAR(1/\hat{\lambda}_1)$ and idem for $\boldsymbol{\hat{\gamma}}$.  Additionally, bootstrap can be used to construct confidence intervals for $\hat{\boldsymbol{\beta_{\text{RC}}}}$. Bootstrap confidence intervals are obtained by sampling the individuals in the internal adjustment set separately from the individuals not included in the internal adjustment set and taking the $(100-\alpha)$ percentiles of the obtained distribution.\\
\textbf{Differential outcome measurement error in univariable analyses.} The variance--covariance matrix for the standard regression estimator $(\boldsymbol{\hat{\beta_{\text{RC}}}},1)$ can be estimated similar to non-differential outcome measurement error defined above (by using the measurement error matrices for differential outcome measurement error). Confidence intervals can then be obtained from equation (\ref{app:eq:waldint}). Bootstrap confidence intervals are obtained by sampling the individuals in the internal adjustment set separately from the individuals not included in the internal adjustment set and taking the $(100-\alpha)$ percentiles of the obtained distribution.

\subsection{Maximum Likelihood for replicates studies}\label{app:sec:variance_ml}
The variance--covariance matrix for the maximum likelihood estimator $\boldsymbol{\hat{\beta}_{\text{MLE}}}$ can be approximated by the multivariate delta method \cite{Bartlett2009LMM}. Denote \newline $\boldsymbol{\zeta}^*=(\delta_0, \boldsymbol{\delta_Z}, \sigma^2_{Y|\boldsymbol{Z}}, \kappa_0, \kappa_Y, \boldsymbol{\kappa_Z}, \sigma^2_{X|Y,\boldsymbol{Z}})$, leaving the $\tau^2$ from $\boldsymbol{\zeta}$ in the main text (see section \ref{sec:meascor:mle}) out as this parameter is not needed for the estimation of $\boldsymbol{\beta}=(\alpha, \phi, \boldsymbol{\gamma})$. A standard result from linear mixed models is that the estimators of fixed parameters are asymptotically uncorrelated with the estimators of the variance component parameters \cite{Bartlett2009LMM}. If one further assumes that the estimators from the linear model of $Y$ given $\boldsymbol{Z}$ are uncorrelated with the parameters estimated in the linear mixed model, it follows for large samples that $\hat{\zeta}^*$ is multivariate normal with mean $\zeta$ and variance covariance matrix $\VAR(\hat{\zeta})$ equal to:
{\scriptsize
\begin{gather*} 
\begin{pmatrix}
\VAR(\hat{\delta_0}) & \COV(\hat{\delta_0}, \boldsymbol{\hat{\delta}_Z}) & 0 & 0 & 0 & 0 & 0 \\
\COV(\boldsymbol{\hat{\delta}_Z}, \hat{\delta_0}) & \VAR(\boldsymbol{\hat{\delta}_Z}) & 0 & 0 & 0 & 0 & 0 \\
0 & 0 & \VAR(\hat{\sigma}^2_{Y|\boldsymbol{Z}}) & 0 & 0 & 0 & 0 \\
0 & 0 & 0 & \VAR(\hat{\kappa}_0) & \COV(\hat{\kappa}_0, \hat{\kappa}_Y) & \COV(\hat{\kappa}_0, \boldsymbol{\hat{\kappa}_Z}) & 0 \\
0 & 0 & 0 & \COV(\hat{\kappa}_Y, \hat{\kappa}_0) & \VAR(\hat{\kappa}_Y) & \COV(\hat{\kappa}_Y, \boldsymbol{\hat{\kappa}_Z}) & 0 \\
0 & 0 & 0 & \COV(\boldsymbol{\hat{\kappa}_Z}, \hat{\kappa}_0) & \COV(\boldsymbol{\hat{\kappa}_Z}, \hat{\kappa}_Y) & \VAR(\boldsymbol{\hat{\kappa}_Z}) & 0 \\
0 & 0 & 0 & 0 & 0 & 0 & \VAR(\hat{\sigma}^2_{X|Y,\boldsymbol{Z}})
\end{pmatrix}
\end{gather*}}

If $g:\mathbb{R}^{5 + 2k} \rightarrow \mathbb{R}^{2 + k}$ is the function that transforms $\boldsymbol{\zeta}^*$ to $\boldsymbol{\beta_{\text{ML}}}=(\alpha_{\text{ML}}, \phi_{\text{ML}}, \boldsymbol{\gamma}_{\text{ML}})$, as defined in the main text, then by the multivariate delta method it follows that in large samples:
\begin{eqnarray*}
\boldsymbol{\hat{\beta}}_{\text{ML}} \sim N(\boldsymbol{\beta}_{\text{ML}},\boldsymbol{J}g\VAR(\hat{\zeta})(\boldsymbol{J}g)'),
\end{eqnarray*}
Where $\boldsymbol{J}$ is the Jacobian matrix of $g$:
\begin{equation*}
    \boldsymbol{Jg}=\begin{pmatrix}
    \frac{\partial \phi}{\partial \delta_0} & \frac{\partial \phi}{\partial \boldsymbol{\delta_Z}} & \frac{\partial \phi}{\partial \sigma^2_{Y|\boldsymbol{Z}}} & \dots & \frac{\partial \phi}{\partial \sigma^2_{X|Y,\boldsymbol{Z}}}\\
    \frac{\partial \alpha}{\partial \delta_0} & \frac{\partial \alpha}{\partial \boldsymbol{\delta_Z}} & \frac{\partial \alpha}{\partial \sigma^2_{Y|\boldsymbol{Z}}} & \dots & \frac{\partial \alpha}{\partial \sigma^2_{X|Y,\boldsymbol{Z}}}\\
    \frac{\partial \boldsymbol{\gamma}}{\partial \delta_0} & \frac{\partial \boldsymbol{\gamma}}{\partial \boldsymbol{\delta_Z}} & \frac{\partial \boldsymbol{\gamma}}{\partial \sigma^2_{Y|\boldsymbol{Z}}} & \dots & \frac{\partial \boldsymbol{\gamma}}{\partial \sigma^2_{X|Y,\boldsymbol{Z}}}
    \end{pmatrix}.
\end{equation*}
 Confidence intervals can then be obtained from equation (\ref{app:eq:waldint}). Bootstrap confidence intervals are obtained by sampling the individuals in the internal adjustment set separately from the individuals not included in the internal adjustment set and taking the $(100-\alpha)$ percentiles of the obtained distribution.
\end{appendix}
\end{document}